\documentclass[aps,pra,superscriptaddress,longbibliography,twocolumn]{revtex4-1}
\usepackage[utf8]{inputenc}
\usepackage{color}
\usepackage{verbatim}
\usepackage{float}
\usepackage{amsmath}
\usepackage{amssymb}
\usepackage{graphicx}
\usepackage{microtype}
\usepackage[unicode=true,
 bookmarks=false,
 breaklinks=false,pdfborder={0 0 1},backref=false,colorlinks=true]
 {hyperref}
\hypersetup{pdftitle={Monitoring Quantum Simulators via Quantum Non-Demolition Couplings to Atomic Clock Qubits},
 pdfauthor={D. V. Vasilyev, A. Grankin, M. A. Baranov, L. M. Sieberer, and P. Zoller},
 linkcolor=blue,citecolor=blue,urlcolor=blue,linkcolor=blue,citecolor=blue,urlcolor=blue}

\makeatletter

\providecommand{\tabularnewline}{\\}

\usepackage{braket}

\definecolor{mygreen}{rgb}{0,0.5,0}
\definecolor{myblue}{rgb}{0,0,0.75}
\definecolor{mymagenta}{cmyk}{0,1,0,0.12}

\makeatother

\begin{document}
\title{Monitoring Quantum Simulators via Quantum Non-Demolition Couplings\\
to Atomic Clock Qubits}
\author{Denis V. Vasilyev}
\thanks{Equal contribution}
\author{Andrey Grankin}
\thanks{Equal contribution}
\author{Mikhail A. Baranov}
\author{Lukas M. Sieberer}
\author{Peter Zoller}
\affiliation{Center for Quantum Physics, University of Innsbruck, 6020 Innsbruck,
Austria}
\affiliation{Institute for Quantum Optics and Quantum Information of the Austrian
Academy of Sciences, 6020 Innsbruck, Austria}
\date{\today}
\begin{abstract}
We discuss monitoring the time evolution of an analog quantum simulator
via a quantum non-demolition (QND) coupling to an auxiliary `clock'
qubit. The QND variable of interest is the `energy' of the quantum
many-body system, represented by the Hamiltonian of the quantum simulator.
We describe a physical implementation of the underlying QND Hamiltonian
for Rydberg atoms trapped in tweezer arrays using laser dressing schemes
for a broad class of spin models. As an application, we discuss a
quantum protocol for measuring the spectral form factor of quantum
many-body systems, where the aim is to identify signatures of ergodic
vs. non-ergodic dynamics, which we illustrate for disordered 1D Heisenberg
and Floquet spin models on Rydberg platforms. Our results also provide
the physical ingredients for running quantum phase estimation protocols
for measurement of energies, and preparation of energy eigenstates
for a specified spectral resolution on an analog quantum simulator. 
\end{abstract}
\maketitle

\section{Introduction}

In the standard scenario of analog quantum simulation, a broad and
tunable class of many-body Hamiltonians of interest is designed based
on the resources provided by a particular physical platform. Examples
in different physical platforms include spin models realized with
Rydberg atoms \citep{Labuhn:2016aa,Norcia2018,Cooper2018,Barredo:2018aa,Guardado-Sanchez2018,Nguyen2018,Orioli2018,Omran2019,DeLeseleuc2019,Cortinas2020,Meinert2020,Pupillo2010,Macri2014,Jau:2016aa,Zeiher:2016aa,Borish2020,Glaetzle:2017aa,Madjarov:2020aa,Signoles2019}
(for review see~\citep{Browaeys2020}), trapped ions~\citep{Garttner:2017aa,Landsman:2019aa,Elben2020}
or superconducting qubits~\citep{Barends:2016aa,Song2019}, or Hubbard
models realized with bosonic and fermionic atoms in optical lattices~\citep{Parsons1253,Boll1257,Gross2017}.
The physical realization of these Hamiltonians then allows the study
of equilibrium and non-equilibrium phenomena, where the quantities
of interest characterizing the quantum many-body state are spin- or
site-occupation correlation functions. These are inferred from (destructive)
site-resolved readout of spins, or in the case of Hubbard models from
quantum gas microscopy.

\begin{figure*}
\includegraphics[width=1\textwidth]{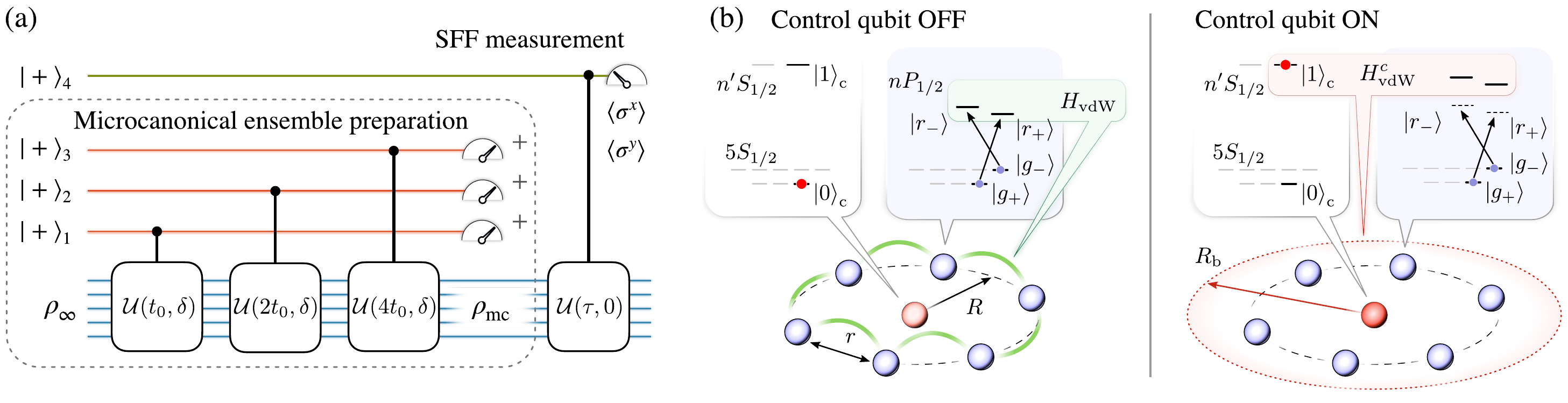}\caption{(a)~Quantum circuit employing a QND coupling of the quantum simulator
to c-qubits to measure SFF. (Dashed block) The simulator spins (blue)
initialized in the infinite temperature ensemble $\rho_{\infty}$
are entangled with $M=3$ c-qubits (red) prepared in states $\ket+_{c}=(\ket0_{c}+\ket1_{c})/\sqrt{2}$
via the QND gate~$\mathcal{U}(t_{m},\delta)=\exp\left\{ -i[(H_{{\rm spin}}-\delta)\otimes\ket0_{c}\bra0]\,t_{m}\right\} $.
Postselecting measurement results of the c-qubits allows one to project
the spins into a microcanonical state $\rho_{{\rm mc}}$. The simulator
in the state $\rho_{{\rm mc}}$ is once again entangled with the c-qubit
(green) to measure the SFF~(see Sec.~\ref{sec:Measurement-of-SFF-Hamiltonian}).
Alternatively, the protocol can be realized with a single c-qubit
via sequential entanglement and measurement cycles. (b)~Engineering
the QND Hamiltonian~\eqref{eq:QNDHamiltonian} with Rydberg-dressed
atoms. (Left panel) The c-qubit (red) in the state $\ket0_{c}$ does
not affect the evolution of the simulator spins (blue) arranged in
a ring of radius $R$ around the control atom. The spins (separated
by a distance $r$) are represented by the hyperfine ground states
$\ket{g_{\pm}}$ of $^{87}{\rm Rb}$ and interact via virtually excited
Rydberg states $\ket{r_{\pm}}$. (Right panel) The c-qubit excited
to the Rydberg state $\ket1_{c}$ breaks the dressing scheme for the
simulator atoms within the Rydberg blockade radius $R_{{\rm b}}$,
thus, blocking the free evolution and realizing the controlled unitary
$\mathcal{U}_{{\rm QND}}(t)$.}
\label{fig:1}
\end{figure*}

In contrast, we will be interested below in a setting where we learn
the state and dynamics of the many-body system by entangling the state
of the quantum simulator with an auxiliary quantum system, followed
by measurement of the auxiliary degrees of freedom, as illustrated
in Fig.~\ref{fig:1}(a). In its simplest form, this auxiliary quantum
system is a single qubit acting as a `clock'~(as in the clock protocol
\citep{Norcia2019,Madjarov2019,Kaubruegger2019}) or `control' qubit,
which can be manipulated by single qubit operations (rotations) and
which we denote c-qubit below. However, the following considerations
generalize immediately also to a multi-qubit setting, where the auxiliary
system represents a small scale quantum memory or quantum computer.

At the heart of our considerations is the quantum non-demolition (QND)
Hamiltonian
\begin{equation}
{\cal H}_{{\rm QND}}=H_{{\rm spin}}\otimes\ket{0}_{c}\bra{0}\label{eq:QNDHamiltonian}
\end{equation}
which generates the QND gate 
\[
\mathcal{U}_{{\rm QND}}(t)=\exp[-i{\cal H}_{{\rm QND}}t],
\]
entangling the quantum simulator with a c-qubit. To be specific, we
assume a spin model with Hamiltonian $H_{{\rm spin}}$ for the simulator,
and we denote by $\{\ket{0}_{c},\ket{1}_{c}\}$ the logical states
of the c-qubit~\footnote{An alternative definition is ${\cal H}_{{\rm QND}}=H_{{\rm spin}}\otimes\sigma_{c}^{z}$
with Pauli operator $\sigma_{c}^{z}$. However, we prefer the form
~\eqref{eq:QNDHamiltonian} in light of the physical realization
in Sec.~\ref{sec:Making-nondemolition-gate}.}. The above Hamiltonian is QND, with $H_{{\rm spin}}$ the `energy'
of the quantum many-body system, which plays the role of the QND variable.
To illustrate the action of the above QND gate ${\cal U}_{{\rm QND}}(t)$,
consider a quantum simulator prepared in superposition $\ket{\psi_{\textrm{spin}}}=\sum_{\ell}c_{\ell}\ket{\ell}$
of energy eigenstates, $H_{\textrm{spin}}\ket{\ell}=E_{\ell}\ket{\ell}$
with eigenenergies $E_{\ell}$. Under the QND gate, an initial state
$\ket{\Psi(t=0)}=\ket{\psi_{\textrm{spin}}}\otimes\ket{+}_{c}$ of
the joint quantum simulator prepared in $\ket{\psi_{\textrm{spin}}}$
and the c-qubit prepared in the superposition state $\ket{+}_{c}=\frac{1}{\sqrt{2}}\left(\ket{0}_{c}+\ket{1}_{c}\right)$
will evolve according to 

\begin{align*}
\ket{\Psi(t=0)} & \rightarrow\ket{\Psi(t)}={\cal U}_{{\rm QND}}(t)\ket{\Psi(0)}\\
 & \quad=\sum_{\ell}c_{\ell}\ket{\ell}\otimes\frac{1}{\sqrt{2}}\left(e^{-iE_{\ell}t}\ket{0}_{c}+\ket{1}_{c}\right).
\end{align*}
Thus the superposition of many-body energy eigenstates gets entangled
with the phase of the Bloch vector of the c-qubit rotating in the
$xy$-plane on the Bloch sphere. A readout of this phase via the c-qubit
provides us with a QND measurement of `the energy' of the quantum
many-body system. In a broader context, we note that this QND gate
is also the basic building block of quantum algorithms like quantum
phase estimation (QPE)~\citep{Nielsen2011,Giedke2006,Svore2014}
to measure energies of the many-body spin system, and prepare corresponding
eigenstates with a prescribed spectral resolution. We will use such
features below in a protocol to measure the spectral form factor (SFF),
where\textbf{ }we access correlations between eigenenergies $E_{\ell}$,
encoded through a Fourier transform, via the c-qubit.

In the present paper we will first address the challenge of implementing
the QND Hamiltonian~\eqref{eq:QNDHamiltonian} and QND gate for a
broad class of\emph{ }freely designable spin models in a quantum simulator
in atomic physics setups. Remarkably, as we show in Sec.~\ref{sec:Making-nondemolition-gate},
this can be achieved with Rydberg tweezer platforms~\citep{Labuhn:2016aa,Norcia2018,Cooper2018,Barredo:2018aa,Guardado-Sanchez2018,Omran2019,DeLeseleuc2019}
using laser dressing schemes~\citep{Pupillo2010,Macri2014,Glaetzle:2017aa,Jau:2016aa,Zeiher:2016aa,Borish2020},
and by employing a Rydberg blockade mechanism between the c-qubit
and the simulator atoms to implement the QND Hamiltonian~\eqref{eq:QNDHamiltonian}.
Second, we wish to explore and illustrate the application of quantum
protocols, which build on the above QND gate, providing access to
novel quantum many-body observables of interest under experimentally
realistic conditions. A relevant example is provided by measurement
of the SFF, as discussed in context of~\citep{Cotler2017a,Kos2018,Suntajs2019,Abanin2019},
where the aim is to identify signatures of ergodic vs. non-ergodic
dynamics without an explicit spectroscopic study~\citep{Senko2014,Jurcevic2015,Roushan2017}
of the energy spectrum. In Sec.~\ref{sec:Measurement-of-SFF-Hamiltonian}
we describe a protocol, where the SFF can be measured via QND couplings
to a c-qubit. We illustrate this protocol and its performance with
simulated measurement runs for the disordered 1D Heisenberg model
and Floquet models, which can be implemented with our techniques on
Rydberg platforms.

The paper is organized as follows. In Sec.~\ref{sec:Making-nondemolition-gate}
we discuss implementation of the QND Hamiltonian with Rydberg tweezer
arrays. The properties of the SFF in Hamiltonian systems are briefly
summarized in Sec.~\ref{sec:Measurement-of-SFF-Hamiltonian}, where
we also present the protocol for preparation of the initial state
(Sec.~\ref{subsec:Preparation MCE}) and the protocol for the SFF
measurement together with the discussion of its experimental limitations
(Sec.~\ref{subsec:SFF_protocol_in Hamiltonian systems}). The SFF
of Floquet systems and the corresponding measurement protocol are
considered in Sec.~\ref{sec:Measurement-of-SFF-Floquet}, and we
conclude in Sec.~\ref{sec:Conclusions-and-Outlook}.

\section{Physical Realization of ${\cal H}_{{\rm QND}}$ in Rydberg tweezer
arrays\label{sec:Making-nondemolition-gate}}

The challenge is to implement ${\cal H}_{{\rm QND}}$, Eq.~\eqref{eq:QNDHamiltonian},
for a broad and tunable class of spin models $H_{{\rm spin}}$ on
a given physical platform. The relevant example below is the disordered
1D Heisenberg spin-$1/2$ model,

\begin{equation}
H_{{\rm spin}}=\sum_{i<j=1}^{L}\sum_{\eta=x,y,z}J_{ij}^{\left(\eta\right)}\sigma_{i}^{\eta}\sigma_{j}^{\eta}+\sum_{i=1}^{L}h_{i}^{z}\sigma_{i}^{z}\,,\label{eq:Hspin}
\end{equation}
or, focusing on a specific choice of the couplings used below we have
\begin{align}
H_{\text{spin}} & =J\sum_{i=1}^{L}\Big\{(\sigma_{i}^{x}\sigma_{i+1}^{x}+\sigma_{i}^{y}\sigma_{i+1}^{y}+\Delta\sigma_{i}^{z}\sigma_{i+1}^{z})\nonumber \\
 & +J_{2}(\sigma_{i}^{x}\sigma_{i+2}^{x}+\sigma_{i}^{y}\sigma_{i+2}^{y})+\Delta_{2}\sigma_{i}^{z}\sigma_{i+2}^{z}\Big\}\label{eq:Heasinberg_model}\\
 & +\sum_{i=1}^{L}h_{i}^{z}\sigma_{i}^{z}.\nonumber 
\end{align}
with designable \emph{single} particle and \emph{two-body} (interaction)
terms. In implementing ${\cal H}_{{\rm QND}}$ we are required to
implement the \emph{two} and \emph{three-body} terms involving the
c-qubit as ${\cal H}_{{\rm QND}}=H_{{\rm spin}}\otimes\ket{0}_{c}\bra{0}$.

Rydberg tweezer arrays in 1D, 2D and 3D~\citep{Labuhn:2016aa,Norcia2018,Cooper2018,Barredo:2018aa,Guardado-Sanchez2018,Omran2019,DeLeseleuc2019}
provide the tools to design a broad class of spin Hamiltonians $H_{{\rm spin}}$
via Rydberg dressing~\citep{Pupillo2010,Macri2014,Glaetzle:2017aa,Zeiher:2016aa,Jau:2016aa,Borish2020}.
Here long-lived atomic (hyperfine) ground states are trapped in the
optical tweezer, and play the role of the spins in our quantum simulator,
while the two-body interactions are engineered by admixing weakly
to the ground state via off-resonant laser dressing the (strong) van
der Waals interactions between Rydberg states~\citep{GDN15,Bijnen2015}.
Remarkably, the \emph{same} `dressing toolbox' which allows to design
$H_{{\rm spin}}$ also provides us with a recipe to engineer ${\cal H}_{{\rm QND}}$. 

Engineering of two-qubit entangling gates using Rydberg blockade is
a well established field~\citep{Jaksch2000,SWM10,Hankin2014,Ravets:2014aa,Maller2015,Madjarov:2020aa}.
Here we are interested in a multi-qubit QND gate where (as originally
discussed in \citep{MLW09}, see also~\citep{PZS16,Zhu2016,Grusdt2016,Serbyn2017,Xu2019,Young2020})
interactions engineered via laser dressing can be turned on and off
by preparing a c-qubit in the ground state $\ket{0}_{c}$ or a Rydberg
state $\ket{1}_{c}$, respectively {[}see Fig.~\ref{fig:1}(b) left
and right panel, respectively{]}. If the c-qubit is in the ground
state, it does not interact with the system spins, and thus the Hamiltonian
$H_{{\rm spin}}$ is realized, as discussed above. On the other hand,
for the c-qubit in the Rydberg state $\ket{1}_{c}$, the long-range
character and strength of Rydberg-Rydberg interactions will, via the
\emph{dipole blockade mechanism}, detune the Rydberg states of the
simulator atoms, thus effectively turning off the dressing interactions,
i.e. we have ${\cal H}_{{\rm QND}}=H_{{\rm spin}}\otimes\ket{0}_{c}\bra{0}$.
We will analyze this below in a realistic atomic physics setting.

\subsection{Hamiltonian for the simulator and c-qubit\label{subsec:Atomic-configuration} }

We consider the setup outlined in Fig.~\ref{fig:1}(b). $L$ atoms
trapped in optical tweezers (with trapping frequency $\omega_{{\rm trap}}$)
are arranged in a ring representing a 1D quantum simulator of spin-$1/2$
with periodic boundary conditions. The distance $r$ between simulator
atoms is assumed to be larger than $2.4\,\mu\text{m}$ as discussed
in Appendix~\eqref{subsec:Validity-of-the}. The simulator atoms
are assumed to be close to their motional ground state. We also consider
the effective spin dynamics to be adiabatic with respect to the atomic
motion, $J_{ij}\ll\omega_{{\rm trap}}$, and thus neglect the latter.
This condition is easily satisfied in the considered setup.

The c-qubit is represented by an atom trapped in the center of the
ring (more c-qubits can be realized with atoms placed, for example,
on the axis of the ring). In Fig.~\ref{fig:1}(b) we show the atomic
level structure for the atoms representing the quantum simulator,
and the control atom. The spin-1/2 degrees of freedom of the simulator
are encoded in two long-lived hyperfine ground states $\left|g_{\pm}\right\rangle $
with energies $E_{g\pm}$. These are coupled to Rydberg states $\left|g_{\pm}\right\rangle \rightarrow\left|r_{\pm}\right\rangle $
with energies $E_{r\pm}$ by two off-resonant lasers with respective
frequencies $\omega_{\pm}$ and detunings $\Delta_{\pm}=\omega_{\pm}-E_{r\pm}+E_{g\pm}\gg\Omega_{\pm}$,
and $\Omega_{\pm}$ corresponding Rabi frequencies. The ground state
of the c-qubit is $\left|0\right\rangle _{\text{c}}$ with energy
$E_{gc}$, while $\left|1\right\rangle _{\text{c}}$ is a Rydberg
state with energy energy $E_{rc}$. We drive the control atom transition
with a laser of frequency $\omega_{c}$ which is tuned near resonance
with detuning $\Delta_{c}=\omega_{c}-E_{cr}+E_{gc}$. The corresponding
Rabi frequency is $\Omega_{c}$. In our protocol, the c-qubit is prepared
initially in a superposition state of ground and excited state with
a $\pi/2$-pulse.

The Hamiltonian of the total system is $H_{\textrm{tot}}=H_{s}+H_{c}+H_{sc}$,
written as sum of the simulator and control atom Hamiltonians, and
the simulator-control interaction. These are given by\begin{widetext}
\begin{align}
H_{s} & =\left\{ \sum_{i=1}^{L}\sum_{\alpha=\pm}\left[E_{g_{\alpha}}\ket{g_{\alpha}}_{i}\bra{g_{\alpha}}+E_{r_{\alpha}}\ket{r_{\alpha}}_{i}\bra{r_{\alpha}}+(\Omega_{\alpha}\ket{g_{\alpha}}_{i}\bra{r_{\alpha}}e^{-i\omega_{\alpha}t}+\text{H.c.})\right]+\sum_{i<j}^{L}H_{\text{vdW}}^{\left(i,j\right)}\right\} \otimes\mathbb{I}_{c},\label{eq:H_s}\\
H_{c} & =\mathbb{I}_{s}\otimes[E_{gc}\ket{0}_{c}\bra{0}+E_{cr}\ket{1}_{c}\bra{1}+(\Omega_{c}\ket{0}_{c}\bra{1}e^{-i\omega_{c}t}+\text{H.c.})],\\
H_{sc} & =\sum_{i=1}^{L}H_{\text{vdW}}^{\left(i,c\right)}.\label{eq:H_sc}
\end{align}
\end{widetext}Here $H_{\text{vdW}}^{\left(i,j\right)}$ is the van
der Waals interaction between the Rydberg manifolds $\{\ket{r_{\pm}}_{i}\}$
and $\{\ket{r_{\pm}}_{j}\}$ of simulator atoms $i$ and $j$, and
$H_{\text{vdW}}^{\left(i,c\right)}$ denotes the van der Waals interaction
between the control atom in $\ket{1}_{c}$ and Rydberg atom $\{\ket{r_{\pm}}_{i}\}$. 

As stated above, in our model, the effective spin-$1/2$ of the quantum
simulator is encoded in ground states $\ket{g_{\pm}}$ of the simulator
atoms (see, for example,~\citep{GDN15} and~\citep{BYB19} for a
specific choice). Following~\citep{GDN15}, an effective spin-spin
interaction is obtained by admixing to these ground states with a
laser a fine-structure split Rydberg state. We note that it is the
combination of a \emph{fine-structure} resolved Rydberg manifold (i.e.
the spin-orbit coupling) together with the van der Waals interaction
(away from Förster resonances~\citep{Ravets:2014aa}) which provides
the physical mechanism for the effective spin-spin interactions in
the ground state manifold, and determines the spin models which can
be realized in this setup (see Appendices~\ref{sec:Appendix.Dipole-dipole-interactions}
and~\ref{sec:Appendix.Adiabatic-elimination-of}). To be specific,
we assume below $^{87}$Rb with Rydberg states~$\ket{r_{\pm}}=\ket{nP_{1/2},m_{j}=\pm1/2}$
for the simulator atoms, and Rydberg state of the control atom~$\ket{n^{\prime}S_{1/2},m_{j}=1/2}$.
We furthermore assume $\left|n-n^{\prime}\right|\gg1$ in order to
avoid direct dipolar exchange interaction between control and simulator
atoms. In the basis $\{\ket{r_{\pm}}_{i}\otimes\ket{r_{\pm}}_{j}\}$
of the given Rydberg manifold, the interaction $H_{\text{vdW}}^{\left(i,j\right)}$
is then represented by a $4\times4$~matrix~\citep{BYB19},
\begin{align}
H_{\text{vdW}}^{\left(i,j\right)}=\frac{1}{r_{ij}^{6}} & \left[C_{6}\mathbb{I}_{4}-\widetilde{C}_{6}\mathbb{D}_{0}\left(\theta_{ij},\phi_{ij}\right)\right].\label{eq:V_vdw-2}
\end{align}
Here $C_{6}$ and $\widetilde{C}_{6}$ are van der Waals interaction
constants, $\mathbb{I}_{4}$ is the identity matrix, and $\mathbb{D}_{0}$
a $4\times4$~matrix with $\theta_{ij},\phi_{ij}$ angles representing
the axis connecting the pair of atoms $i,j$ with distance $r_{ij}$.
Explicit expressions for these quantities are provided in Appendix~\ref{p-p_interaction}.
In a similar way, the van der Waals interaction between the simulator
atom $i$ and control atom $c$ has the form
\begin{align}
H_{\text{vdW}}^{\left(i,c\right)}\approx & \frac{C_{6}^{\prime}}{r_{ic}^{6}}\sum_{\alpha=\pm}\left|r_{\alpha}\right\rangle _{i}\left\langle r_{\alpha}\right|\otimes\ket{1}_{c}\bra{1},\label{eq:vdw_control-1}
\end{align}
with $C_{6}^{\prime}$ the corresponding van der Waals coefficient.
Here we made the assumption that $C_{6}^{\prime}$ is independent
of the state $\ket{r_{\pm}}$ (see Appendix~\ref{p-s_interaction}).
Thus, conditional to the control atom to be in the Rydberg state $\ket{1}_{c}$,
the Rydberg energies of the simulator atoms are shifted by $E_{r\pm}\rightarrow E_{r\pm}+C_{6}^{\prime}/r_{ic}^{6}$,
which for the ring geometry of Fig.~\ref{fig:1}(b) provides an identical
shift for all simulator atoms. In the spirit of the Rydberg blockade
mechanism we assume this shift to be large. Thus, for a control atom
in the ground state $\ket{0}_{c}$ the dressing lasers $\omega_{\pm}$
are detuned from the Rydberg states $\ket{r_{\pm}}_{i}$ by $\Delta_{\pm}$,
while the presence of a control atom in $\ket{1}_{c}$ will detune
these excited states by $\Delta_{\pm}-C_{6}^{\prime}/r_{ic}^{6}$.
For a control atom in $\ket{0}_{c}$ the dressing lasers will thus
induce interactions between the effective ground state spin-$1/2$
by admixing the Rydberg-Rydberg interactions $H_{\text{vdW}}^{\left(i,j\right)}$
to the ground state manifold, while a control atom in $\ket{1}_{c}$
detunes the Rydberg states, and thus effectively shuts off Rydberg
dressing.

\begin{figure}
\begin{centering}
\includegraphics[width=1\columnwidth]{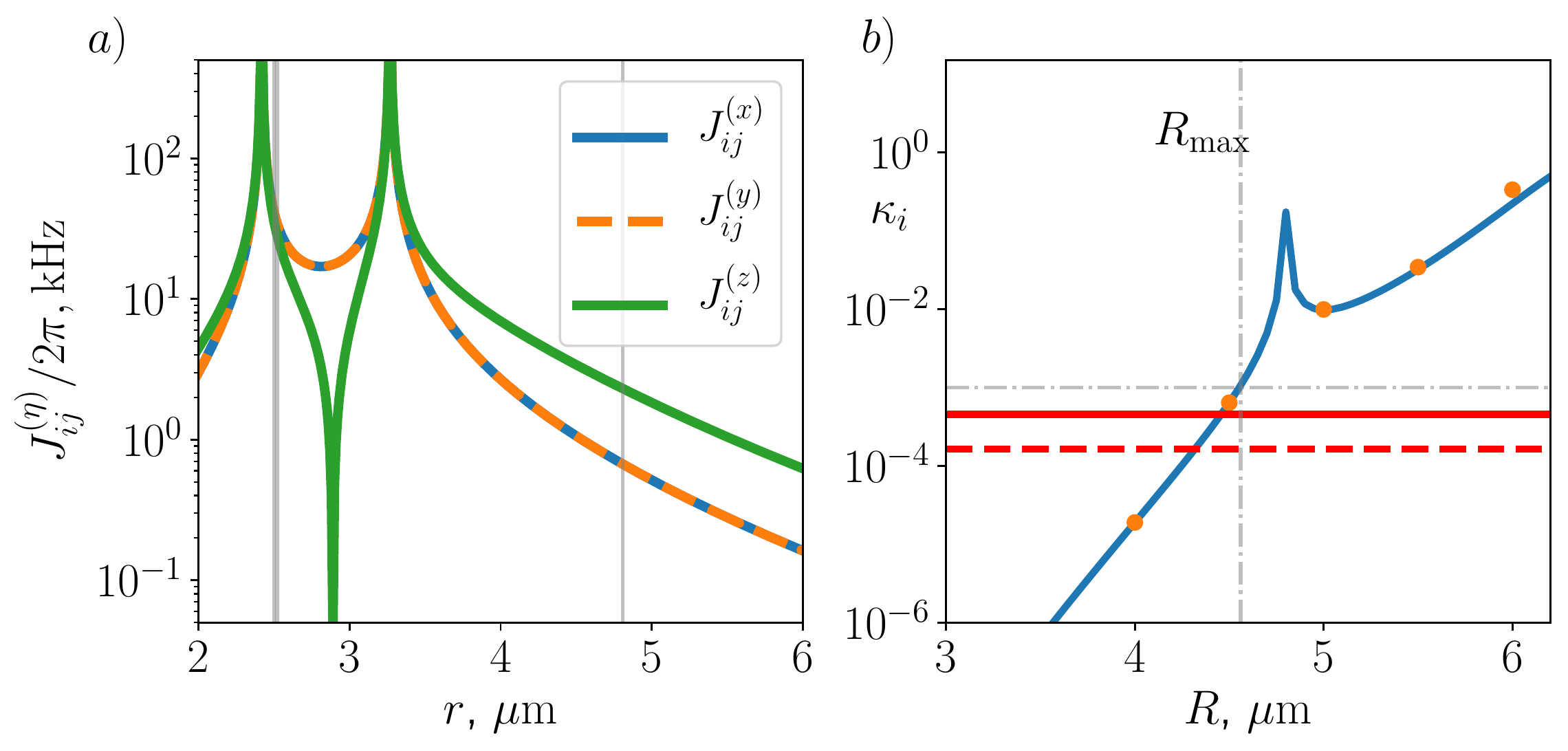}
\par\end{centering}
\caption{Engineered QND Hamiltonian. (a) Spin-spin interaction constants as
a function of interatomic distance $r$. Gray vertical lines indicate
the nearest and next-nearest neighbor positions. (b) Decoherence parameters
$\kappa_{1}$ (red solid), $\kappa_{2}$ (red dashed) and $\kappa_{3}$
(analytic blue solid and numeric orange dots) for $L\sim10$ spins.
Parameters of the dressing are: $\xi_{\pm}\equiv\Omega_{\pm}/\Delta_{\pm}=0.2$,
$\Delta_{\pm}=-9\,\text{MHz}$. Principle quantum number of Rydberg
atoms are $n=60$, $n^{\prime}=71$.}

\label{Fig2}
\end{figure}

\subsection{QND Hamiltonian and imperfections\label{subsec:QND-Hamiltonian}}

The derivation of ${\cal H}_{{\rm QND}}$ proceeds now by a perturbative
elimination of the Rydberg manifold of the simulator atoms. First,
for the control atom in the ground state $\ket{0}_{c}$, and $\Omega_{c}=0$,
the dressing lasers for the simulator atoms are described by the detunings
$\Delta_{\pm}$, as discussed above, and the relevant perturbation
parameter for elimination of the Rydberg states is $\xi_{\pm}=\Omega_{\pm}/\Delta_{\pm}\ll1$.
This yields an effective dynamics as spin-$1/2$ model for the dressed
ground states $\ket{\tilde{g}_{\pm}}_{i}$ with effective spin-$1/2$
Hamiltonian denoted by $H_{{\rm spin}}$. The design of the desired
spin Hamiltonians via dressing schemes by an appropriate choice of
the atomic and laser configurations, and van der Waals interaction
for a given geometry was discussed by \citep{GDN15}. In Appendix~\ref{sec:Appendix.Adiabatic-elimination-of}
we provide a derivation for the disordered 1D Heisenberg chain in
the form~\eqref{eq:Heasinberg_model}. There we list explicit expressions
for the $J^{\left(x,y,z\right)}$, $h$ etc. in Eqs.~(\ref{eq:hz}-\ref{eq:beta_ij})
of Appendix~\ref{sec:Appendix.Adiabatic-elimination-of} as functions
of the microscopic atomic parameters. Figure~\ref{Fig2}(a) shows
the values of the spin-spin interaction constants $J_{ij}^{\left(\eta\right)}$
as a function of the interatomic distance for a particular choice
of dressing scheme parameters.

Second, we repeat this derivation for the control atom in the excited
state $\ket{1}_{c}$ which amounts to replacing the detunings $\Delta_{\pm}\rightarrow\Delta_{\pm}-C_{6}^{\prime}/r_{ic}^{6}$
in the expressions for the parameters of the Hamiltonian \eqref{eq:JxJyJz}
of Appendix~\ref{sec:Appendix.Adiabatic-elimination-of}. This results
in a spin Hamiltonian which we denote by $H'_{{\rm spin}}$, with
the same structure as $H_{{\rm spin}}$, but strongly suppressed couplings
$J'\ll J$ etc. After combining the two cases, we obtain the Hamiltonian
describing the coupling between the simulator and the c-qubit
\[
\tilde{{\cal H}}=H_{\text{spin}}\otimes\ket{0}_{c}\bra{0}+H_{\text{spin}}^{\prime}\otimes\ket{1}_{c}\bra{1}.
\]

If $H_{{\rm spin}}$ and $H_{{\rm spin}}^{\prime}$ commute, $\tilde{{\cal H}}$
is equivalent to the QND Hamiltonian \eqref{eq:QNDHamiltonian} with
$H_{\text{spin}}\to H_{\text{spin}}-H_{\text{spin}}^{\prime}$. In
the opposite case, $H_{\text{spin}}^{\prime}$ results in errors in
the entangling gate $\mathcal{U}_{{\rm QND}}(t)$, with the error
rate characterized by a dimensionless parameter $\kappa_{3}=|H_{{\rm spin}}^{\prime}|/|H_{{\rm spin}}|$,
where $|\ldots|$ denotes the difference between the maximal and the
minimal eigenvalues of the corresponding operator. Another source
of error is related to spontaneous emission of Rydberg states of the
control and simulator atoms, with error rates characterized by $\kappa_{1}\equiv\gamma_{\text{d}}^{\prime}/|H_{\text{spin}}|$
and $\kappa_{2}\equiv\xi_{\pm}^{2}\gamma_{\text{d}}L/|H_{\text{spin}}|$,
respectively, where $\gamma_{\text{d}}^{\prime}$ and $\gamma_{\text{d}}$
are the corresponding spontaneous emission rates {[}see Fig.~\ref{Fig2}(b)
and Appendix~\ref{sec:Appendix.Experimental-considerations} for
more details{]}. The largest error rate determines the coherence time
$t_{\mathrm{coh}}\sim[|H_{\text{spin}}|\max\{\kappa_{i}\}]^{-1}$
below which the errors in the gate operation due to spontaneous emission
and $H_{{\rm spin}}^{\prime}$ can be ignored.

Thus for $t<t_{\mathrm{coh}}$ the quantum simulator coupled to the
c-qubit can be described by the QND Hamiltonian Eq.~\eqref{eq:QNDHamiltonian}
\begin{equation}
\widetilde{{\cal H}}_{\mathrm{QND}}=\left(H_{\text{spin}}-\delta\right)\otimes\ket{0}_{c}\bra{0},\label{eq:H_system-qubit-1}
\end{equation}
which is written here in the rotating frame with respect to the laser
of the c-qubit, and we have set $\Omega_{c}=0$ in $H_{c}$ ($\Omega_{c}\neq0$
is needed only during the preparation stage of the c-qubit and for
the measurement readout). In the above equation $\delta$ refers to
a renormalized detuning of the control laser (see Appendix~\ref{sec:Appendix.Adiabatic-elimination-of}).

The geometric constraints on the possible configurations of atoms
in space are imposed by the validity conditions of the van der Waals
interaction Hamiltonians (\ref{eq:V_vdw-2}, \ref{eq:vdw_control-1})
and by the efficiency of the Rydberg blockade. In particular, the
distances between control and simulator atoms~$R$ and between simulator
atoms~$r$ are limited to $R<R_{{\rm b}}\approx6.5\mu\text{m}$ and
$r\gtrsim2.4\mu\text{m}$, respectively (see Appendix~\ref{subsec:Validity-of-the}
and \ref{sec:Appendix.Experimental-considerations}). This restricts
the number of atoms in 1D ring geometry to 12. Larger system sizes
can be achieved in 2D or quasi-2D arrangements of atoms. For example,
the number of atoms can be doubled by considering two co-axial rings
of atoms. We note that in this case the resulting spin model parameters
are not described by Fig.~\ref{Fig2}(a) due to a more complex van
der Waals interaction pattern for inter-ring pairs of atoms. Alternatively,
atoms can form a small 2D lattice fitting into the blockade radius
$R_{{\rm b}}$ of the control atom placed above the lattice.

\section{Measurement of the SFF in Hamiltonian Systems\label{sec:Measurement-of-SFF-Hamiltonian}}

For a generic quantum many-body system, the energy level statistics
is an indicator which allows to distinguish between quantum ergodic
and non-ergodic regimes~\citep{Haake2010,Pal2010,Luitz2015,DAlessio2016}.
Quantum ergodic (or chaotic) systems are characterized by eigenenergies
statistics given by the Wigner-Dyson distribution~\citep{Mehta2004}
in a universality class of random matrices, and the eigenfunctions
are delocalized over the configuration space. In contrast, non-ergodic
quantum systems, such as integrable models, display Poisson statistics
of energy levels and localized wave-functions. These considerations
apply not only to systems with time-independent Hamiltonians and their
energy spectrum, but also to time-periodic Floquet systems~\citep{Haake2010}
characterized by quasienergies.

\begin{figure}
\includegraphics[width=0.9\columnwidth]{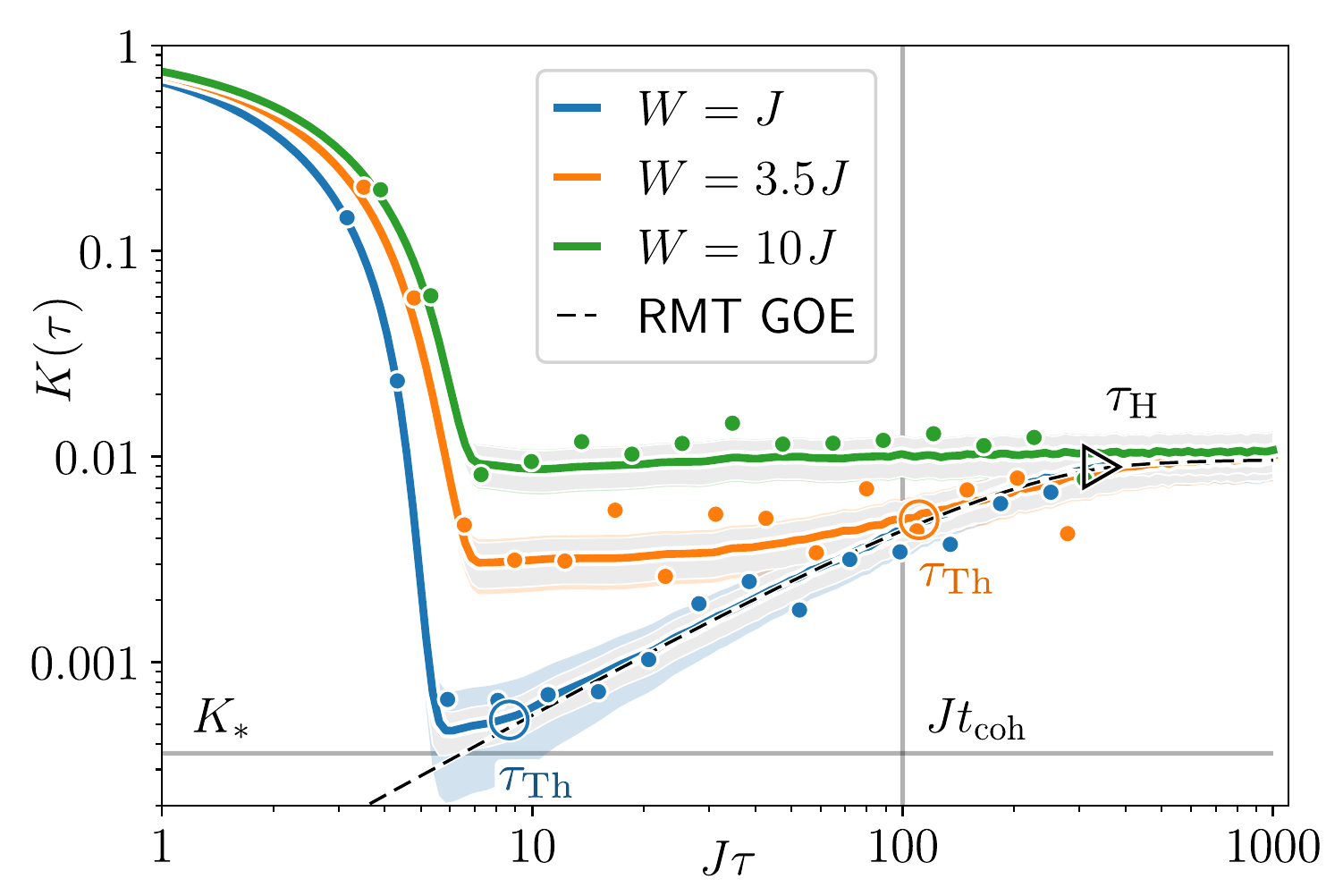}

\caption{Measurement of the spectral form factor $K(\tau)$ in the Heisenberg
chain \eqref{eq:Heasinberg_model} of $L=12$ spins for different
disorder strengths $W$. The black dashed line represents the RMT
prediction $K_{{\rm GOE}}(\tau)$~\eqref{eq:KGOE}. The color dots
show the results of the simulated measurements of $K(\tau)$. The
color solid lines and the shaded areas indicate, respectively, the
numerical prediction for $K(\tau)$ and the root-mean-square error
due to the shot noise and disorder averaging for the simulated $\simeq2\times10^{5}$
experimental runs per data point. The gray areas show the contribution
of averaging over 20 disorder realizations. The color circles indicate
the Thouless times $\tau_{{\rm Th}}$ and the black triangle shows
the Heisenberg time $\tau_{{\rm H}}$. The parameters of the Heisenberg
model~\eqref{eq:Heasinberg_model} are $\Delta=0.8$, $J_{2}=0.02$,
$\Delta_{2}=0.06$.}
\label{fig:SSF_L=00003D12}
\end{figure}

An equivalent characterization of ergodic vs.$\:$non-ergodic dynamics
is provided by the SFF, where spectral features are displayed in the
time domain, i.e.$\:$essentially as the Fourier transform of the
two-point correlation function of the spectral density. For a many-body
Hamiltonian $H_{{\rm spin}}$ with eigenenergies $E_{\ell}$ the SFF
is defined as
\begin{align}
K(\tau) & =\overline{\big|\sum_{\ell}f(E_{\ell})e^{-iE_{\ell}\tau}\big|^{2}}\label{eq: definition SFF-1}\\
 & \equiv\overline{\big|\mathrm{tr}(e^{-iH_{\textrm{spin}}\tau}\rho_{\mathrm{mc}})\big|^{2}}.
\end{align}
Here $f(E)$ is a nonnegative normalized smooth filter function of
width $\Delta E$ covering a band in the middle of the energy spectrum,
where the density of states is flat, thus probing the properties of
typical states of the many-body system, while eliminating contributions
from spectral edges. The overline indicates a possible disorder average.
The second line of (\ref{eq: definition SFF-1}) rewrites the SFF
in terms of the microcanonical density operator $\rho_{\mathrm{mc}}=\sum_{\ell}f(E_{\ell})\ket{\ell}\bra{\ell}$
representing an initial density matrix. The SFF has been central in
high-energy physics~\citep{Cotler2017a} and the recent discussion
of the transition from many-body localization (MBL) to quantum chaos
in a class of generic spin chains with disorder~\citep{Kos2018,Suntajs2019,Abanin2019}. 

\subsection{Properties of the SFF}

To illustrate the generic behavior of SFF, we show in Fig.~\ref{fig:SSF_L=00003D12}
a numerically calculated SFF for the disordered Heisenberg XXZ spin-1/2
chain of length $L=12$ with Hamiltonian \eqref{eq:Heasinberg_model}.
A random local transverse field {[}last term in Eq.~\eqref{eq:Heasinberg_model}{]}
plays the role of disorder, with values $h_{i}\in[-W,W]$ drawn from
a uniform distribution and $W$ characterizing the disorder strength.
We assume periodic boundary conditions, and consider the sector with
zero total spin projection on the $z$-axis, $S_{z}=\sum_{i=1}^{L}\sigma_{i}^{z}=0$.
The choice of this model is based on the following considerations.
First, it exhibits a transition from quantum chaos for weak disorder
to many-body localization for strong disorder, and second, it belongs
to the class of models, where the QND-gate and thus the SFF protocol
of the following section can be implemented with the dressing scheme
in Rydberg tweezer arrays (see Sec.~\ref{sec:Making-nondemolition-gate}). 

According to Fig.~\ref{fig:SSF_L=00003D12}, for small times~$\tau$
the SFF $K(\tau)$ decays from its initial value $K(0)=1$ due to
dephasing on a time scale $\sim1/\Delta E$. Here $\Delta E\approx(E_{\mathrm{max}}-E_{\mathrm{min}})/6$,
i.e., $1/6$ of the width of the energy spectrum. The signature of
quantum chaos is the existence of a ramp at long times~$\tau$: Random
Matrix Theory (RMT) for the Gaussian orthogonal ensemble (GOE), which
is applicable to systems obeying time-reversal symmetry, predicts

\begin{equation}
K_{{\rm GOE}}(\tau)=K_{\infty}\begin{cases}
2\frac{\tau}{\tau_{{\rm H}}}-\frac{\tau}{\tau_{{\rm H}}}\log(1+2\frac{\tau}{\tau_{{\rm H}}}), & 0<\tau\leq\tau_{{\rm H}},\\
2-\frac{\tau}{\tau_{{\rm H}}}\log\left(\frac{2\tau+\tau_{{\rm H}}}{2\tau-\tau_{{\rm H}}}\right), & \tau>\tau_{{\rm H}},
\end{cases}\label{eq:KGOE}
\end{equation}
which is shown as dashed line in Fig.~\ref{fig:SSF_L=00003D12}.
In the formula above, $\tau_{{\rm H}}=2\pi/\delta_{E}\sim2^{L}/L$
denotes the Heisenberg time associated with the mean level spacing
$\delta_{E}=\left\langle E_{\ell+1}-E_{\ell}\right\rangle $ in the
middle of the spectrum. Furthermore, $K_{\infty}=\sum_{\ell}f^{2}(E_{\ell})$,
where the value of $K_{\infty}$ is equivalently given by the inverse
of the number $N_{\Delta E}$ of the eigenstates in the energy interval
$\Delta E$, $K_{\infty}\approx N_{\Delta E}^{-1}$. As shown in Fig.~\ref{fig:SSF_L=00003D12},
the RMT prediction agrees well with simulations for a finite size
chain in the limit of weak disorder (blue line) for times $\tau>\tau_{\mathrm{Th}}$,
with $\tau_{\mathrm{Th}}$ the Thouless time defined as the onset
of the ramp~\citep{Suntajs2019}. The flattening with increasing
disorder strength (orange and green lines) is indicative of the crossover
towards nonergodic (MBL) behavior. 

The Hamiltonian~\eqref{eq:Heasinberg_model} is time-reversal symmetric
and thus its chaotic phase is described by the GOE. As discussed in
Appendix~\ref{subsec:Complex J}, this model can also be realized
with complex $J$, so that the chaotic phase is described by a GUE.
In systems with broken time-reversal symmetry, $K(\tau)$ follows
the RMT prediction for the Gaussian unitary ensemble (GUE),

\begin{equation}
K_{{\rm GUE}}(\tau)=K_{\infty}\begin{cases}
\frac{\tau}{\tau_{{\rm H}}}, & 0<\tau\leq\tau_{{\rm H}},\\
1, & \tau>\tau_{{\rm H}}.
\end{cases}\label{eq:KGUE}
\end{equation}

Our SFF protocol below exploits repeated QND measurements of the c-qubit
to both prepare the desired initial energy distribution $f(E)$ in
\eqref{eq: definition SFF-1} {[}see Fig.~\ref{fig:1}(a){]}, as
well as to read the SFF \eqref{eq: definition SFF-1}. Challenges
faced in measuring the SFF, and addressed below, are decoherence times
in quantum simulators, relative to times required to identify the
`ramp', and the number of measurements required to provide clear signatures
of both the ergodic and non-ergodic regimes. Figure~\ref{fig:SSF_L=00003D12}
plots simulated measurements for a finite measurement budget which
compare favorably with the (exact) numerical results for the SFF (solid
lines). The question to be addressed is whether, with given experimental
resources, it is possible to see the main characteristics of the SFF.

\subsection{Measurement Protocol via QND-Coupling to a c-qubit}

An experimental protocol to measure the SFF in the Hamiltonian case
will require: (i) the ability to prepare an initial (microcanonical)
ensemble of states $\rho_{\mathrm{mc}}=\sum_{\ell}f(E_{\ell})\ket{\ell}\bra{\ell}$
with a filter function $f(E)$ of given width $\Delta E$ in the center
of the spectrum; (ii) the ability to resolve for a given number of
measurements, and thus signal-to-noise ratio the baseline $K_{\infty}\approx N_{\Delta E}^{-1}$
and the value of the SFF around the Thouless time; and finally (iii)
the ability to observe for a given decoherence time (part of) the
ramp $\tau_{{\rm Th}}<\tau<\tau_{{\rm H}}$ and possibly the long
time behavior of $K(\tau)$. 

\subsubsection{Preparation (verification) of a microcanonical ensemble\label{subsec:Preparation MCE}}

The first step of the protocol requires preparation of the microcanonical
(MC) ensemble $\rho_{{\rm mc}}=\sum_{\ell}f(E_{\ell})\ket{\ell}\bra{\ell}$.
This can be achieved with a low resolution phase estimation algorithm
(PEA), providing a probabilistic preparation of an energy band via
repeated measurement of the c-qubit. The PEA is based on the QND Hamiltonian
\eqref{eq:QNDHamiltonian}, and requires a minimal number of measurements
$M$ of the c-qubit to achieve a given measurement precision $\Delta E$~\citep{Nielsen2011,Giedke2006,Svore2014}.

\begin{figure}
\includegraphics[width=1\columnwidth]{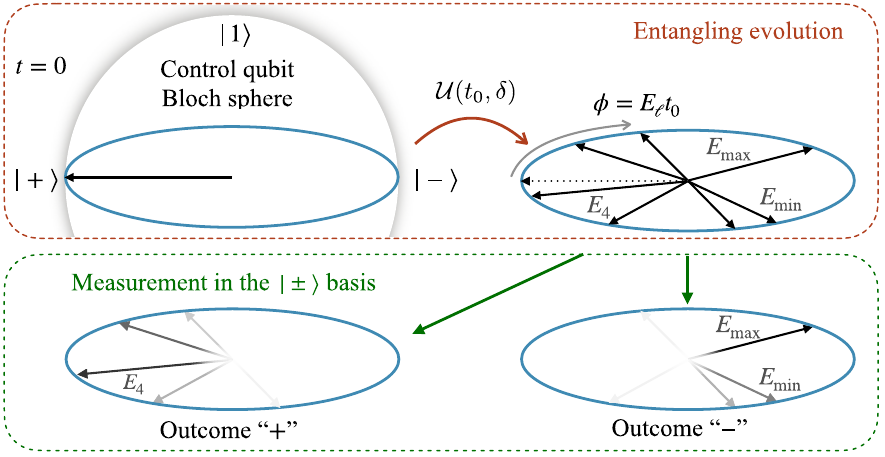}

\caption{The microcanonical ensemble preparation process. The first filtering
step with interaction time $t_{0}$ is shown. The QND interaction~\eqref{eq:QNDHamiltonian}
rotates the c-qubit proportionally to the eigenenergy $E_{\ell}$
of the spin Hamiltonian~\eqref{eq:Heasinberg_model}. The measurement
in the basis $\ket\pm$ leads to narrowing of the populated energy
window of the spins state $\rho_{{\rm out}}$. The shades of gray
of the arrows corresponding to different $E_{\ell}$ represent the
conditional probability for the eigenstate $\ket\ell$ to appear in
$\rho_{{\rm out}}$.}
\label{fig:MC_1step}
\end{figure}

Figure~\ref{fig:MC_1step} illustrates the idea of the preparation
protocol. First, the many-body spin system initialized in a state
$\rho_{{\rm in}}$ and the c-qubit prepared in the state $\ket{+}_{c}$
are entangled by the QND interaction $\mathcal{H}_{\textrm{QND}}$~\eqref{eq:QNDHamiltonian}
during a time $t_{0}$. As a result the c-qubit is rotated proportionally
to the values of eigenenergies $E_{\ell}$ of the spin Hamiltonian~\eqref{eq:Heasinberg_model}.
The time $t_{0}$ is chosen to maximally spread the full spectrum
$E_{\ell}$ of the Hamiltonian $H_{{\rm spin}}$ over the equator
of the c-qubit Bloch sphere. After that, the c-qubit measurement in
$\ket\pm_{c}$ basis shrinks the populated energy window of the spins
state by a factor of $\sim2$. Repeating the cycle with increasing
interaction times $t_{m}$ and postselecting ``$+$'' outcomes results
in the state $\rho_{{\rm mc}}$ with a narrow energy distribution.

\begin{figure}
\includegraphics[width=0.9\columnwidth]{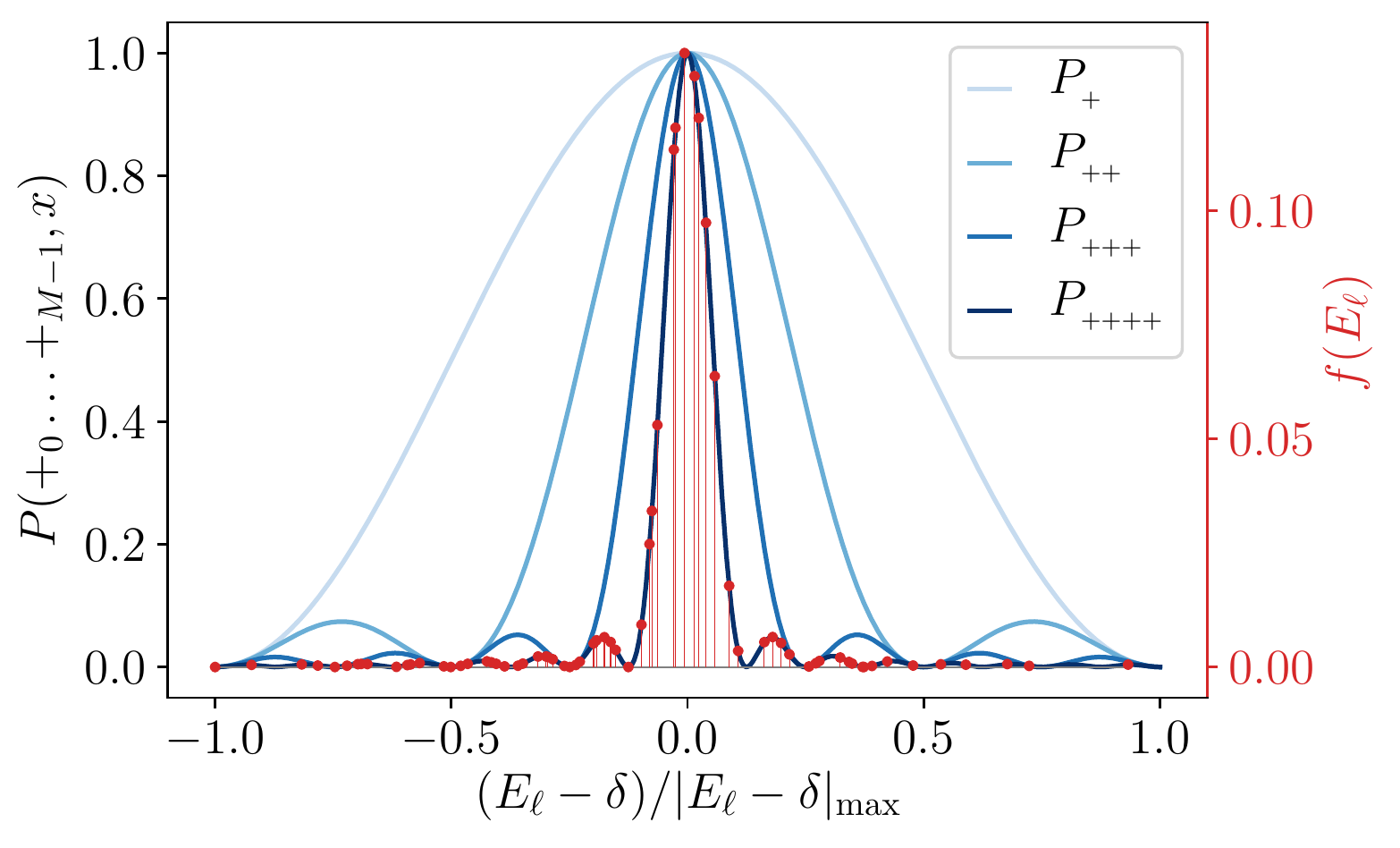}\caption{The filter functions for $M=1,2,3,4$ filtering steps (c-qubits) are
shown in blue. The eigenstates probabilities $f(E_{\ell})$ filtered
with $M=4$ steps are shown in red.}
\label{fig:MC_preparation}
\end{figure}

Figure~\ref{fig:1}(a) shows the quantum circuit for the full preparation
procedure involving $M=3$ c-qubits entangled with the simulator via
\[
\mathcal{U}(t_{m},\delta)=\exp\left\{ -i[(H_{{\rm spin}}-\delta)\otimes\ket0_{c}\bra0]\,t_{m}\right\} .
\]
Note, that the same result can be achieved by performing sequential
entanglement and measurement cycles with a single c-qubit. The interaction
times are chosen as $t_{m}\equiv2^{m}t_{0},\,m=0,\ldots,M-1$ where
$t_{0}\leq\pi/(|E_{\ell}-\delta|_{{\rm max}})$. The renormalized
detuning of the control laser $\delta$ introduced in Eq.~\eqref{eq:H_system-qubit-1}
allows to tune the energy band of the state to be prepared. If we
now select a run with all readouts ``$+_{m}$'', $m=0,\ldots,M-1$,
then the initial state of the spin system is projected into the state
$\rho_{{\rm out}}$ with a narrow distribution in the energy eigenbasis
$\ket\ell$ (see Appendix~\ref{sec:Appendix.Preparation-of-MCE})
\[
\bra\ell\rho_{{\rm out}}\ket\ell\propto p_{\ell}=P(+_{0}\ldots+_{M-1},E_{\ell}-\delta)\,\bra\ell\rho_{{\rm in}}\ket\ell,
\]
where
\begin{equation}
P(+_{0}\ldots+_{M-1},x)=\left\{ \frac{\sin(2^{M}t_{0}x)}{2^{M}\sin(t_{0}x)}\right\} ^{2}.\label{eq:mc_filter}
\end{equation}
The success probability of the preparation is $p_{{\rm mc}}\equiv\sum_{\ell}p_{\ell}$.
The function~\eqref{eq:mc_filter} has a peak at $x=0$ with a width
$\sim2^{-M}/t_{0}$, and, therefore, the conditional outcome state
exhibits an exponential narrowing of the energy distribution/uncertainty
around $E_{\ell}\approx\delta$. Note, however, that the variance
of this distribution has the same scaling $\sim2^{-M}/t_{0}^{2}$
due to the presence of long tails.

Following the above procedure, the microcanonical ensemble can be
prepared by initializing the spin system in the infinite temperature
state $\rho_{{\rm in}}=\rho_{\infty}\propto\sum_{\ell}\ket\ell\bra\ell$
and postselecting the outcome state with the probability $p_{{\rm mc}}\sim2^{-M}$.
As a result, the spin system will be probabilistically prepared in
the microcanonical ensemble $\rho_{{\rm mc}}=\sum_{\ell}f(E_{\ell})\ket\ell\bra\ell$
with $f(E_{\ell})\propto\sum_{\ell}P(+_{0}\ldots+_{M-1},E_{\ell}-\delta)$.
The state has the mean energy $\bar{E}={\rm Tr}\{H_{{\rm spin}}\rho_{{\rm mc}}\}=\delta$
and the bandwidth $\Delta E=\sqrt{{\rm Tr}\{(H_{{\rm spin}}-\delta)^{2}\rho_{{\rm mc}}\}}\sim2^{-M/2}/t_{0}$.
Note that the initial infinite temperature ensemble $\rho_{\infty}$
can be sampled by random initialization of individual spins in up
and down states where we additionally apply the constraint $S_{z}=0$~(equal
number of spins up and down) to probe the SFF in the zero-magnetization
sector. An example of the resulting eigenstates probability distributions
is shown in Fig.~\ref{fig:MC_preparation} for the Heisenberg chain~\eqref{eq:Heasinberg_model}
of $L=8$ spins.

We remark that measuring the probability $p_{{\rm mc}}$ of the successful
MC ensemble preparation allows one to estimate (see Appendix~\ref{sec:Appendix.Measurement-of-Heisenberg}
for details) the Heisenberg time $\tau_{{\rm H}}$ and the plateau
value $K_{\infty}$ of the SFF as $\tau_{{\rm H}}\simeq2^{M+1}t_{0}\mathcal{D}p_{{\rm mc}}$
and $K_{\infty}\simeq(2/3)(\mathcal{D}p_{{\rm mc}})^{-1}$. Here $\mathcal{D}$
is the Hilbert space dimension of the considered symmetry sector of
the model. For the Heisenberg spin chain~\eqref{eq:Heasinberg_model}
with $S_{z}=0$ we have $\mathcal{D}=C_{L}^{L/2}\approx\sqrt{2/\pi L}2^{L}$
with $C_{L}^{L/2}$ being the binomial coefficient (we assume $L$
to be even). In the experiment the probability $p_{{\rm mc}}$ is
given by the ratio of the number of successful preparation attempts
to the total number of experimental runs. The estimated values of
$\tau_{{\rm H}}$ and $K_{\infty}$ uniquely determine the behavior
of the SFF in the chaotic regime assuming RMT (dashed lines in Figs.~\ref{fig:SSF_L=00003D12}
and \ref{fig:SFF_experiment_simulation}).

The low resolution PEA procedure can also be used to verify that a
given state consists of a superposition of energy eigenstates in a
narrow energy interval $\Delta E$ around $\bar{E}$ (MC ensemble):
With $\delta=\bar{E}$, the appearance of $M\simeq2\log_{2}(|E_{\ell}-\delta|_{{\rm max}}/\Delta E)$
successive ``$+$'' readouts with probability close to $1$ signals
that the given state has a desired energy variance. Further, if the
state $\rho_{{\rm in}}$ contains a collection of excited states in
a narrow energy interval created from some initial state by a time-dependent
perturbation (see \citep{Senko2014}, for example), a MC state can
be distilled with our procedure through several successive ``$+$''
readouts%
.

\subsubsection{Protocol to measure the SFF \label{subsec:SFF_protocol_in Hamiltonian systems}}

Following the preparation of the microcanonical ensemble $\rho_{\textrm{mc}}$,
we perform the evolution for a time $\tau$ with the QND-Hamiltonian
\eqref{eq:QNDHamiltonian}, and finally measure the expectation values
of $\sigma^{x}$ and $\sigma^{y}$ for the c-qubit, as shown in the
Fig.~\ref{fig:1}(a),

\begin{align*}
\braket{\sigma^{x}(\tau)} & ={\rm Tr}\!\left\{ \!e^{-i\tau H_{{\rm spin}}\otimes\ket{0}_{c}\!\bra{0}}\!\rho_{{\rm mc}}\!\otimes\!\ket{+\!}_{c}\!\bra{\!+}\!e^{i\tau H_{{\rm spin}}\otimes\ket{0}_{c}\!\bra{0}}\sigma^{x}\!\right\} \\
 & =\sum_{\ell}f(E_{\ell})\cos(E_{\ell}\tau),\\
\braket{\sigma^{y}(\tau)} & =\sum_{\ell}f(E_{\ell})\sin(E_{\ell}\tau),
\end{align*}
These provide us with $|\sum_{\ell}f(E_{\ell})e^{-iE_{\ell}\tau}\big|^{2}\approx\braket{\sigma^{x}(\tau)}^{2}+\braket{\sigma^{y}(\tau)}^{2}$
(see Appendix~\ref{sec:Appendix.Measurement noise}). We obtain $K(\tau)$
by repeating this sequence for different disorder realizations and
averaging the result.

The SFF measurement scheme realizes a QND measurement meaning that
the initial state of the spin system is not heated up or destroyed
after the interaction with the c-qubit. It is, therefore, possible
to reuse the once prepared $\rho_{{\rm mc}}$ for the measurement
of $K(\tau)$ with different times $\tau_{i}$ (see Appendix~\ref{sec:Appendix.Recycling-of-the}).
The maximum number of recycling times $N_{{\rm reuse}}$ is restricted
by the coherence time $t_{{\rm coh}}$ of the spin system as $\sum_{i=1}^{N_{{\rm reuse}}}\tau_{i}\ll t_{{\rm coh}}$.

\subsubsection{Experimental challenges\label{subsec:Experimental-challenges}}

An experimental realization of the SFF measurement faces two main
challenges, which limit the achievable system sizes.

\emph{Time scales:} – Propagation up to the Heisenberg time $\tau_{{\rm H}}$,
which grows exponentially with the system size $L$, is limited by
the finite coherence time of the quantum simulator. Thus, observation
of the behavior of $K(\tau)$ at times $\tau\sim\tau_{\mathrm{H}}$,
requires coherence times $t_{\mathrm{coh}}>\tau_{\mathrm{H}}\sim2^{L}/(JL)$.
We note, however, that the effects of interest such as the transition
to chaotic dynamics at the Thouless time $\tau_{{\rm Th}}$ and the
distinct behaviors of the SFF for quantum chaotic {[}$K(\tau)\sim\tau${]}
and integrable systems {[}$K(\tau)\sim{\rm const.}${]} take place
at much shorter times to be compared with $t_{{\rm coh}}$ (see Appendix~\ref{sec:Appendix.Experimental-considerations}).

\emph{Signal magnitude:} – The second challenge is the exponentially
small values of the SFF $\sim2^{-L}$ at the characteristic times.
The threshold signal level which can be distinguished from the shot
noise after averaging over $N_{{\rm d}}$ realizations of disorder
in the spin Hamiltonian and $N$ measurements per one disorder realization
is given by $K_{*}\equiv2(1+\sqrt{2})/(N\sqrt{N_{{\rm d}}})$ (see
Appendix~\ref{sec:Appendix.Measurement noise}). Thus, the total
number of experimental runs per data point necessary to resolve the
features of interest in $K(\tau)$, is given by $N_{{\rm run}}>2^{L}\sqrt{N_{{\rm d}}}/N_{{\rm reuse}}$
for the probabilistic preparation of the initial MC ensemble (see
Appendix~\ref{sec:Appendix.Experimental-considerations}).

\subsubsection{Numerical simulation of the SFF measurement\label{subsec:Numerical-simulation-of-SFF}}

To demonstrate the feasibility of the SFF measurement with our protocol,
we perform numerical simulations of the measurement process in the
disordered Heisenberg spin chain for various numbers $L$ of spins.
The simulation includes the probabilistic preparation of the initial
state $\rho_{{\rm mc}}$ and averaging over finite numbers of disorder
realizations and experimental runs.

\begin{figure}
\includegraphics[width=0.9\columnwidth]{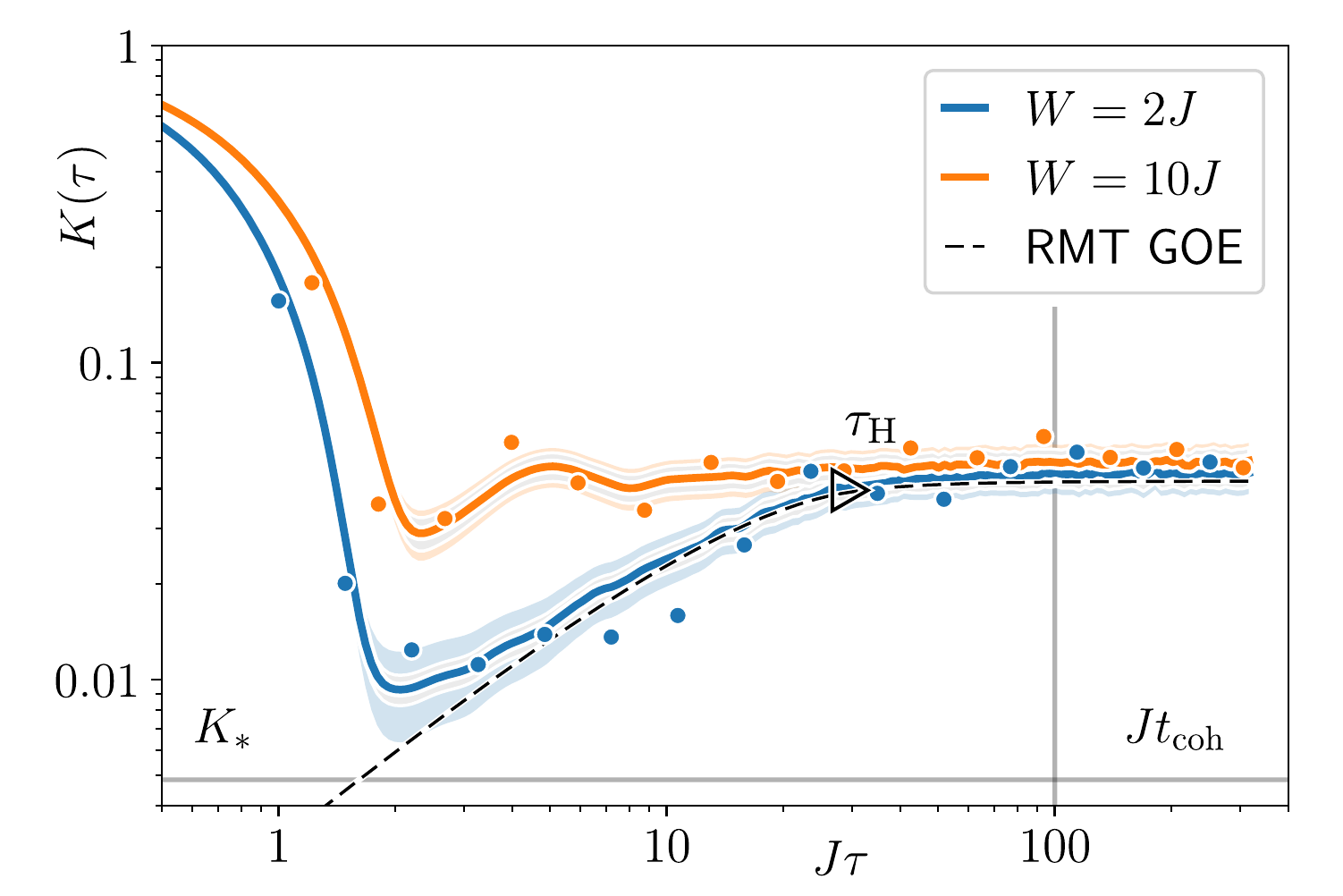}\caption{Revealing the signatures of quantum chaos in the disordered Heisenberg
chain of $L=8$ spins. The colored dots show results of simulated
measurements of $K(\tau)$ for two disorder strengths $W=2J,10J$.
Solid lines show the corresponding theoretical $K(\tau)$ and the
black dashed line represents the RMT prediction $K_{{\rm GOE}}(\tau)$.
The shaded areas show the root-mean-square error due to the shot noise
and disorder averaging. The black triangle shows the Heisenberg time
$\tau_{{\rm H}}$. The model parameters are the same as in Fig.~\ref{fig:SSF_L=00003D12}
(see also text).}
\label{fig:SFF_experiment_simulation}
\end{figure}

The results are presented in Figs.~\ref{fig:SSF_L=00003D12} and
\ref{fig:SFF_experiment_simulation}. According to Fig.~\ref{fig:SFF_experiment_simulation},
a small system of $L=8$ spins with the measurement budget of $N_{{\rm run}}\simeq10^{4}$
experimental runs per data point allows identification of the linear
ramp in $K(\tau)$. This indicates the chaotic behavior with level
repulsion in the Heisenberg model with a weak disorder $W=2J$ (blue
dots). In contrast, a strong disorder $W=10J$ results in the localized
behavior of the system dynamics (orange dots) lacking correlations
in the distribution of the eigenenergies. The colored lines and the
corresponding shaded areas represent the numerical prediction for
$K(\tau)$ and the root-mean-square error of the simulated measurement,
respectively. The black dashed line is the RMT prediction $K_{{\rm GOE}}(\tau)$
Eq.~\eqref{eq:KGOE}.

The Thouless times $\tau_{{\rm Th}}$ for various disorder strengths
can be probed in a larger system of $L=12$ spins as shown in Fig.~\ref{fig:SSF_L=00003D12}.
It is evident from the figure that the time $\tau$ at which the data
points approach the RMT prediction $K_{{\rm GOE}}(\tau)$ given by
the black dashed line grows with the increase of the disorder strength
$W$, which is compatible with the expected behavior of the Thouless
time $\tau_{{\rm Th}}$. The root-mean-square error of the simulated
measurement with $N_{{\rm run}}\simeq2\times10^{5}$ experimental
runs per data point is shown with shaded areas around the solid lines
indicating the numerical prediction for $K(\tau)$.

In both Figs.~\ref{fig:SSF_L=00003D12} and \ref{fig:SFF_experiment_simulation}
the horizontal and vertical gray lines mark the shot noise threshold
$K_{*}$ and the coherence time $t_{{\rm coh}}\sim10^{2}J^{-1}$ in
our Rydberg setup (see Appendix~\ref{sec:Appendix.Experimental-considerations}),
respectively. For the system of $L=8(12)$ spins we simulate MC ensemble
preparation with $M=3(5)$ filtering steps, perform averaging over
$N_{{\rm d}}=100(20)$, and include recycling of the prepared MC ensemble
for $N_{{\rm reuse}}=10$ times. The parameters of the Heisenberg
model~\eqref{eq:Heasinberg_model} for both system sizes $\Delta=0.8$,
$J_{2}=0.02$, $\Delta_{2}=0.06$.

It is important to stress, that in both cases of $L=8$ and $L=12$,
the $K_{{\rm GOE}}(\tau)$ curves (black dashed lines) are completely
determined by the plateau value $K_{\infty}$ for $\tau\to\infty$
and by the Heisenberg time $\tau_{{\rm H}}$, both of which can be
independently estimated as outlined in Sec.~\ref{subsec:Preparation MCE}
above and discussed in Appendix~\ref{sec:Appendix.Measurement-of-Heisenberg}. 

\section{Measurement of the SFF in Floquet Systems\label{sec:Measurement-of-SFF-Floquet}}

To study aspects of ergodicity, thermalization, and quantum chaotic
dynamics, periodically driven or Floquet systems are particularly
appealing for several reasons: First, due to the absence of energy
conservation, generic Floquet systems can thermalize very rapidly
and completely even for relatively small system sizes \citep{Zhang2016}.
Further, since the density of quasienergies is generically flat, it
is not necessary to unfold the spectra of Floquet systems to access
their spectral statistics \citep{Haake2010}. This relative simplicity
of Floquet systems has led to intriguing recent results which, through
explicit calculations of the SFF, establish an analytical connection
between RMT and many-body quantum chaos in interacting periodically
driven spin chains \citep{Kos2018,Bertini2018,Chan2018,Chan2018b}.
As we discuss below, such Floquet spin models, and the experimental
measurement of the SFF, can be realized naturally with Rydberg dressing
schemes.

\subsection{SFF of Floquet systems}

In Floquet systems, we are interested in the statistics of eigenvalues
of the unitary operator $U(\vartheta)$ which describes the evolution
of the system during one driving period of duration $\vartheta$:
$U(\vartheta)\ket\ell=e^{-i\lambda_{\ell}\vartheta}\ket\ell$ with
$\lambda_{l}\in[0,\,2\pi/\vartheta]$ being the quasienergy. Similar
to the Hamiltonian case {[}see Eq.~\eqref{eq: definition SFF-1}{]},
we define the SFF for Floquet systems as~\citep{Bertini2018}
\begin{equation}
K(t)=\overline{\big|\sum_{\ell}f_{l}e^{-i\lambda_{\ell}\vartheta t}\big|^{2}}=\overline{|\mathrm{tr}[U(\vartheta)^{t}\rho_{\mathrm{in}}]|^{2}},\label{eq:SFF-FLoquet}
\end{equation}
where the integer ``time'' $t$ is the number of the evolution periods,
$\rho_{\mathrm{in}}=\sum_{\ell}f_{l}\ket{\ell}\bra{\ell}$ is the
initial quantum state, and the overline represents a possible average
over disorder. The flat density of states of quantum chaotic Floquet
systems allows one to use the infinite temperature ensemble $\rho_{\mathrm{in}}=\rho_{\infty}=\mathcal{D}^{-1}\sum_{\ell}\ket{\ell}\bra{\ell}\propto\mathbb{I}$
as the initial state (here $\mathcal{D}$ is the dimension of the
Hilbert space).

The behavior of the SFF in Floquet dynamics has the same characteristic
features as in the Hamiltonian case. As an illustration, below we
consider two Floquet systems. First, we present a Floquet model which
exhibits a crossover between the circular orthogonal ensemble (COE)
and the circular unitary one (CUE) as a function of the driving frequency
$\omega=2\pi/\vartheta$~\citep{Regnault2016}. Then we consider
kicked Ising models which demonstrate clear signatures of COE and
CUE statistics even for a small system of $L=4$ spins.

\emph{Crossover between COE and CUE statistics}. – Here we consider
an interesting example of a periodically driven system for which a
random matrix class of the Floquet operator $U(\vartheta)$ differs
from that of the time-dependent Hamiltonian $H(t)$ which generates
the dynamics of the system~\citep{Regnault2016}. To be specific,
we consider a piecewise constant Hamiltonian with $H_{1}$ during
the first half of the driving period and $H_{2}$ during the second
half. The evolution operator over one period is of the form

\begin{equation}
U(\vartheta)=e^{-i(\vartheta/2)H_{1}}e^{-i(\vartheta/2)H_{2}},\label{eq:licked_Heisenberg}
\end{equation}
where $H_{1}$ and $H_{2}$ are Heisenberg Hamiltonians with random
magnetic fields $h_{i}^{x,y,z}$, which are normally distributed around
zero with unit variance,

\begin{align*}
H_{1} & =J\sum_{i}^{L}\left\{ \sum_{\eta=x,y,z}\sigma_{i}^{\eta}\sigma_{i+1}^{\eta}+\frac{1}{2}(h_{i}^{x}\sigma_{i}^{x}+h_{i}^{y}\sigma_{i}^{y})\right\} ,\\
H_{2} & =J\sum_{i}^{L}\left\{ \sum_{\eta=x,y,z}\sigma_{i}^{\eta}\sigma_{i+1}^{\eta}+\frac{1}{2}(h_{i}^{z}\sigma_{i}^{z}-h_{i}^{y}\sigma_{i}^{y})\right\} .
\end{align*}
The Heisenberg models can be realized in the quantum simulator based
on dressed Rydberg atoms as discussed in Sec.~\ref{sec:Making-nondemolition-gate}.
For generic values of the driving period $\vartheta$ the spectral
statistics of the Floquet operator $U(\vartheta)$ is described by
the CUE. However, for finite system sizes and in the limit of high
driving frequencies, the dynamics of the system is described by an
effective time-independent Hamiltonian $(H_{1}+H_{2})/2$ in which
the random fields in the $y$ direction cancel and time-reversal symmetry
is restored. Therefore, in this limit, the spectral statistics of
$U(\vartheta)$ belongs to the COE. We note that in this limit the
density of quasienergies is determined by the effective Hamiltonian
and is not flat. However, in the numerical examples below we find
that effects due to a nonflat density of quasienergies are insignificant,
and we take $\rho_{\mathrm{in}}$ in Eq.~\eqref{eq:SFF-FLoquet}
to be the infinite temperature ensemble also for high driving frequencies.

\begin{figure}
\includegraphics[width=0.9\columnwidth]{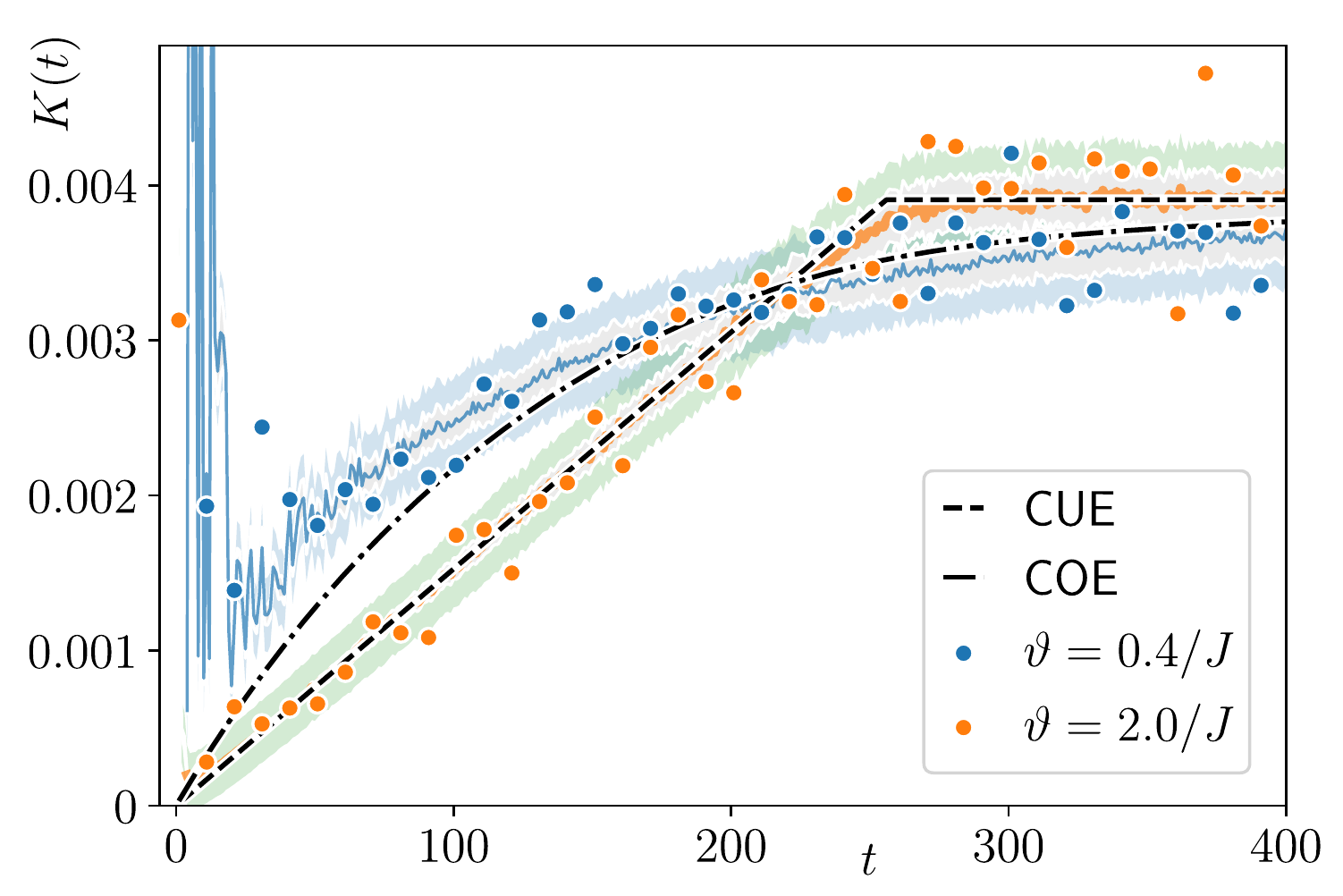}\caption{Spectral form factor in the disordered Floquet model Eq.~\eqref{eq:licked_Heisenberg}.
The model exhibits a crossover between COE (blue dots) and CUE (orange
dots) statistics as a function of the driving period $\vartheta$.
The black dashed and dot-dashed lines show the RMT predictions $K_{{\rm CUE}}(t)$
and $K_{{\rm COE}}(t)$, respectively. Numerical simulations are performed
for $L=8$ spins with $5\times10^{5}$ measurements per data point.}
\label{fig:SFF_floquet}
\end{figure}
 
\begin{figure}
\includegraphics[width=0.9\columnwidth]{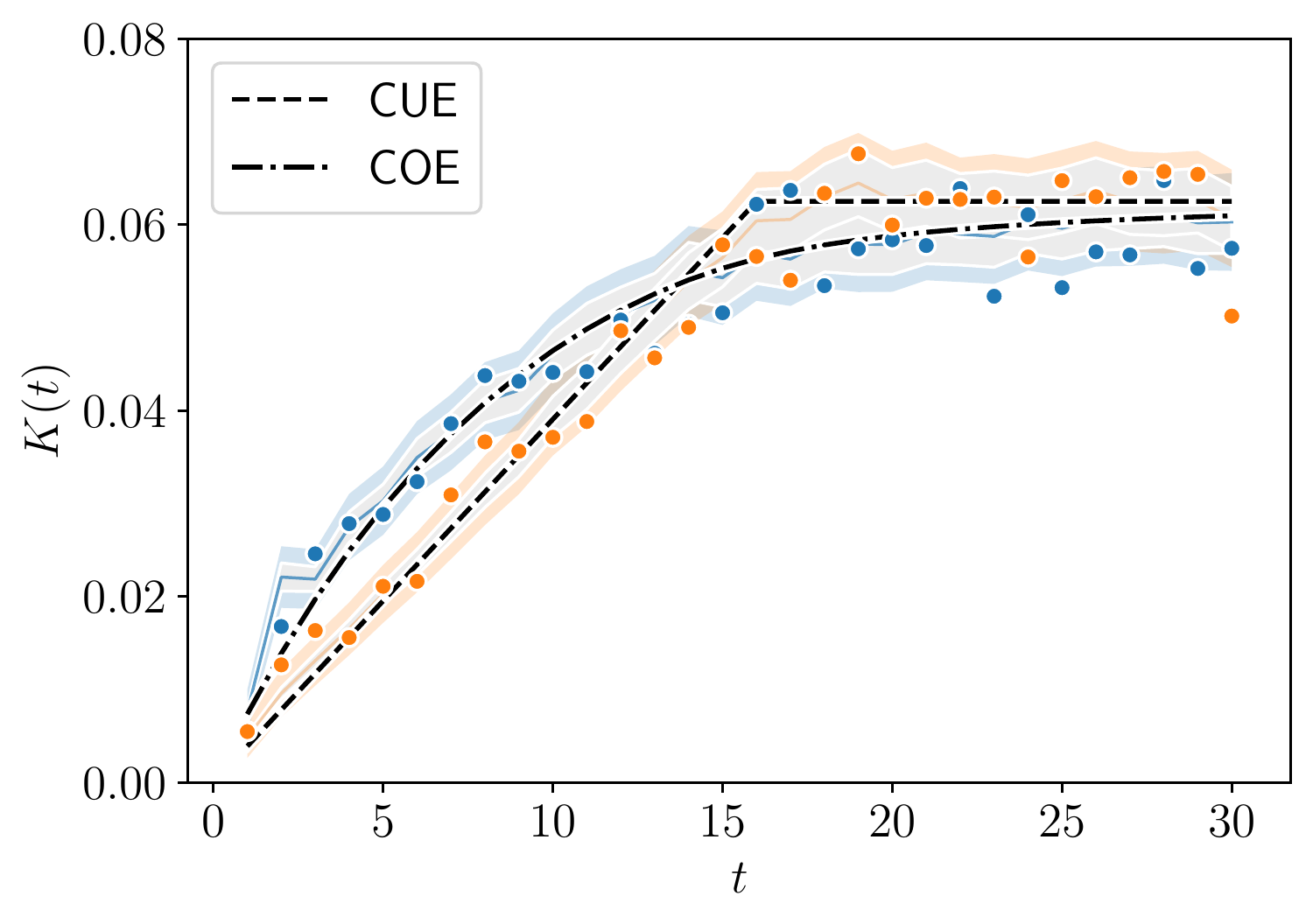}\caption{Spectral form factor in the disordered kicked Ising models Eqs.~\eqref{eq:kicked_Ising_2},~\eqref{eq:kicked_Ising_3}.
Already in the small system of $L=4$ spins the models exhibits a
clear distinction between COE {[}$U_{2}(\vartheta)$ model shown with
blue dots{]} and CUE {[}$U_{3}(\vartheta)$ model shown with orange
dots{]} statistics. The black dashed and dot-dashed lines show the
RMT predictions $K_{{\rm CUE}}(t)$ and $K_{{\rm COE}}(t)$, respectively.
Numerical simulations are performed for $3.6\times10^{4}$ measurements
per data point with driving period $\vartheta=1/J$.}
\label{fig:SFF_floquet_Ising}
\end{figure}

The SFF and the simulated measurement of the SFF (see Sec.~\ref{subsec:Measurement-Protocol-Floquet}
below) are shown in Fig.~\eqref{fig:SFF_floquet}. Remarkably, even
the relatively small Floquet system of $L=8$ spins exhibits clear
RMT behavior with a crossover between COE and CUE statistics upon
changing the Floquet period $\vartheta$. In particular, we observe
COE statistics for short periods or high driving frequencies (blue
dots) with $\vartheta<\vartheta_{{\rm c}}\sim0.5/J$. The dot-dashed
line shows the corresponding RMT prediction $K_{{\rm COE}}(t)=\left[2t-t\ln(1+2t/\mathcal{D})\right]/\mathcal{D}^{2}$
for $0<t<\mathcal{D}$ where the Heisenberg time is set by the Hilbert
space dimension $\mathcal{D}=2^{L}$, see Eq.~\eqref{eq:KGOE}. For
long periods $\vartheta>\vartheta_{{\rm c}}$, the system is in the
unitary class with CUE statistics (orange dots) resulting in $K_{{\rm CUE}}(t)=t/\mathcal{D}^{2}$
for $0<t<\mathcal{D}$ (dashed line), see also Eq.~\eqref{eq:KGUE}.
We note that for sufficiently large values of $\vartheta$ the initial
decay due to dephasing lasts at most a few driving cycles. This is
because for these values of $\vartheta$ the unitary $U(\vartheta)$
is not a sparse matrix, as it is for $\vartheta\ll1$, but a dense
one coupling practically all states in the Hilbert space with each
other. As a result, chaotic behavior starts already at $t=t_{\mathrm{Th}}\sim1$.

\emph{Small Floquet systems}. – Now we consider two kicked Ising models
described by evolution operators 
\begin{align}
U_{2}(\vartheta) & =e^{-iH_{x}\vartheta}e^{-iH_{y}\vartheta},\label{eq:kicked_Ising_2}\\
U_{3}(\vartheta) & =e^{-iH_{x}\vartheta}e^{-iH_{y}\vartheta}e^{-iH_{z}\vartheta},\label{eq:kicked_Ising_3}
\end{align}
where $H_{x,y,z}$ are the transverse Ising Hamiltonians with random
magnetic fields $h_{i}^{x,y,z}\in[-1,1]$
\begin{align*}
H_{x} & =J\sum_{i}^{L}\left(\sigma_{i}^{x}\sigma_{i+1}^{x}+h_{i}^{y}\sigma_{i}^{y}\right),\\
H_{y} & =J\sum_{i}^{L}\left(\sigma_{i}^{y}\sigma_{i+1}^{y}+h_{i}^{z}\sigma_{i}^{z}\right),\\
H_{z} & =J\sum_{i}^{L}\left(\sigma_{i}^{z}\sigma_{i+1}^{z}+h_{i}^{x}\sigma_{i}^{x}\right).
\end{align*}
The two models described by the Floquet operators $U_{2}(\vartheta)$
and $U_{3}(\vartheta)$ belong to the COE and CUE random matrix classes,
respectively. Remarkably, the statistical distinction can already
be seen in a small system of $L=4$ spins as the simulated measurement
of the SFF shows in Fig.~\ref{fig:SFF_floquet_Ising}. In the quantum
simulator based on Rydberg atoms the Ising models can be realized
according to the general scheme presented in the Sec.~\ref{sec:Making-nondemolition-gate}. 

These examples illustrate that, as in the case of the Hamiltonian
systems, the SFF provides a sensitive tool for probing quantum chaotic
behavior of Floquet systems. At the same time, the measurement of
the SFF in Floquet systems faces the same challenges as are present
in Hamiltonian systems, i.e., small values of the signal and exponentially
long time scales. However, for experimental studies of many-body quantum
chaos, Floquet systems can be beneficial because they typically exhibit
pronounced RMT behavior even for comparatively small system sizes.
Moreover, since spectral filtering is not required in Floquet systems,
the preparation step described in Sec.~\ref{subsec:Preparation MCE}
can be omitted.

\subsection{Measurement Protocol\label{subsec:Measurement-Protocol-Floquet}}

To generalize the measurement protocol for the SFF in Hamiltonian
systems described in Sec.~\ref{sec:Measurement-of-SFF-Hamiltonian}
to Floquet systems, we reformulate it in terms of the unitary operator
$\mathcal{U}(t)$ which describes the coupled evolution of the system
and the c-qubit during $t$ Floquet periods of duration $\vartheta$,

\begin{equation}
\mathcal{U}(t)=U(\vartheta)^{t}\otimes\ket0_{c}\bra0+\mathbb{I}\otimes\ket1_{c}\bra1.\label{eq:control_U}
\end{equation}
For simplicity, we focus here on Floquet systems with piecewise constant
Hamiltonians $H_{k}$ for time periods $\tau_{k-1}<\tau<\tau_{k}$,
where $\tau_{0}=0$ and the time dependence is repeated periodically
with period $\vartheta=\sum_{k}\tau_{k}$. The corresponding Floquet
operator reads $U(\vartheta)=\prod_{k}e^{-i\tau_{k}H_{k}}$. The controlled
evolution Eq.~\eqref{eq:control_U} is achieved by using the QND
interaction Hamiltonian \eqref{eq:QNDHamiltonian} between the spin
system and the c-qubit with $H_{{\rm spin}}\equiv H_{k}$:
\[
\mathcal{U}(t)=\Big[\prod_{k}e^{-i\tau_{k}H_{k}\otimes\ket0_{c}\bra0}\Big]^{t}.
\]

The measurement protocol for the SFF starts with the initialization
of the c-qubit in the state $\ket+$ and the system spins in the infinite
temperature state $\rho_{\infty}$. (In practice, it is sufficient
to sample from the infinite temperature ensemble by preparing the
system, e.g., in random product states.) Then, for a particular realization
of disorder in the instantaneous Hamiltonians $H_{k}$, we apply the
controlled evolution $\mathcal{U}(t)$, and measure the expectation
values of the operators $\sigma_{x}$ and $\sigma_{y}$ for the c-qubit
afterwards,
\begin{align*}
\braket{\sigma^{x}(t)} & ={\rm Tr}\left\{ \mathcal{U}(t)\rho_{{\rm \infty}}\otimes\ket{+}_{c}\bra{+}\mathcal{U}(t)^{\dagger}\sigma^{x}\right\} ,\\
 & ={\rm Re}\braket{U(\vartheta)^{t}}\\
\braket{\sigma^{y}(t)} & ={\rm Im}\braket{U(\vartheta)^{t}}.
\end{align*}
Finally, the quantity $|\braket{U(\vartheta)^{t}}|^{2}=\braket{\sigma^{x}(t)}^{2}+\braket{\sigma^{y}(t)}^{2}$
is averaged over disorders realizations resulting in the SFF $K(t)$
for the Floquet system Eq.~\eqref{eq:SFF-FLoquet}.

Experimental limitations of the measurement of the SFF in Floquet
systems are analogous to the Hamiltonian case as discussed in Sec.~\ref{subsec:SFF_protocol_in Hamiltonian systems}.
As an example, the effect of a finite number of measurements is illustrated
in the Figs.~\ref{fig:SFF_floquet} and~\ref{fig:SFF_floquet_Ising}
for the Floquet system described by Eq.~\eqref{eq:licked_Heisenberg}
and Eqs.~\eqref{eq:kicked_Ising_2}, \eqref{eq:kicked_Ising_3},
respectively.

\section{Conclusions and Outlook\label{sec:Conclusions-and-Outlook}}

Recent experimental studies of ergodicity breaking in quantum many-body
systems focus on the absence of thermalization of local observables
in integrable \citep{Kinoshita2006,Gring2012,Langen2015} and many-body
localized systems \citep{Schreiber2015,Choi2016,Smith2016,Xu2018},
and on the slow growth of entanglement \citep{Lukin2019,Brydges2019,Xu2018}.
While it is firmly established through theoretical work that key signatures
of ergodic vs. nonergodic dynamics are carried by individual eigenstates
of the Hamiltonian of a quantum many-body system and the statistics
of the corresponding eigenvalues, accessing these signatures in experiments
requires novel approaches beyond the standard paradigm of quantum
simulation. As a first main result of the present work, we have developed
a method that enables in-depth experimental studies of the level statistics
of interacting quantum many-body systems through the measurement of
the SSF, and we have discussed the feasibility of observing the SFF
signatures of ergodic (RMT) vs. non-ergodic dynamics for both disordered
1D Heisenberg and Floquet spin models. We conclude that the key features
of RMT in the SFF should be observable for system sizes of ten or
more atoms based on techniques which are available in present, or
will be available in next-generation Rydberg experiments. We also
emphasize that experimental realization will strongly benefit from
the ongoing development to improve coherence times and implement high-fidelity
(non-destructive) read out of the Rydberg qubits, which are aimed
at advancing scalable quantum computing on the Rydberg platform.

In a broader context, our method to measure the SFF is an example
of a quantum protocol in which the state of a quantum simulator is
prepared and monitored via the measurement of an auxiliary qubit,
which is entangled with the quantum simulator by applying a QND gate.
This is the second main result of our work: The implementation of
a QND Hamiltonian ${\cal H}_{\textrm{QND}}=H_{\textrm{spin}}\otimes\ket{0}_{c}\bra{0}$
with Rydberg tweezer arrays, which yields a QND gate as ${\cal U}_{\textrm{QND}}(t)=e^{-i{\cal H}_{\textrm{QND}}t}$.
In the context of a many-body system engineered with a Rydberg tweezer
platform, this QND Hamiltonian can be implemented for a broad class
of spin models specified by a Hamiltonian $H_{\textrm{spin}}$. While
we consider 1D systems in form of a ring with the c-qubit in the center,
our ideas also carry over to more complex 2D simulator geometries~\citep{Labuhn:2016aa,Barredo:2018aa,GDN15}.
In addition, unique opportunities to combine quantum simulation with
atomic clocks are offered by Alkaline Earth atoms~\citep{Daley2008,Mukherjee_2011,Cooper2018,Madjarov:2020aa}. 

An intriguing possibility opened up by the present study is the design
and implementation of more general quantum protocols involving QND
gates entangling the quantum simulator with a freely designable $H_{\textrm{spin}}$
with a set of c-qubits. As noted before, this opens the door to run,
for example, a quantum phase estimation algorithms on analog quantum
simulators. In Ref. \citep{Yang2020}, a continuous readout of `the
energy' of a quantum many-body system was proposed as analog measurement
of a homodyne current, with an implementation for a transverse Ising
model with long range interaction. In contrast, quantum phase estimation
based on the present QND gate with a freely designable $H_{\textrm{spin}}$
provides an essentially universal, digital quantum algorithm to achieve
measurement and preparation of (a narrow band of) energy eigenstates
in an analog quantum simulator setting. Moreover, quantum phase estimation
can be utilized to compute the dynamical response functions of quantum
many-body systems \citep{Sels2019}. This opportunity is of particular
interest in the context of NMR, where the ability to design $H_{\textrm{spin}}$
enables accessing the NMR spectra of molecules which are described
by parametric Heisenberg models \citep{Sels2019_NMR}.

\emph{Note added.} — After submission of the present work, we have
become aware of Ref.~\citep{Young2020} by Young et al., which proposes
to realize multi-qubit Rydberg-blockade gates using microwave-dressed
Rydberg states.
\begin{acknowledgments}
The authors thank M. Lukin for helpful discussions. Work is supported
by the European Union program Horizon 2020 under Grants Agreement
No. 817482 (PASQuanS) and No. 731473 (QuantERA via QTFLAG), the US
Air Force Office of Scientific Research (AFOSR) via IOE Grant No.
FA9550-19-1-7044 LASCEM, by the Simons Collaboration on Ultra-Quantum
Matter, which is a grant from the Simons Foundation (651440, P.Z.),
and by the Institut für Quanteninformation. The parameters for the
Rydberg simulations were obtained using the ARC library~\citep{vsibalic2017arc}.
\end{acknowledgments}

\appendix

\section{Dipole-dipole interactions\label{sec:Appendix.Dipole-dipole-interactions}}

In this Appendix we provide details on the van der Waals interactions
between $nP_{1/2}+nP_{1/2}$ and $nP_{1/2}+n^{\prime}S_{1/2}$ states
as relevant for the model of Sec.~\ref{sec:Making-nondemolition-gate}.
Our discussion adapts and extends Ref.~\citep{GDN15}.

For any pair of atoms $i$ and $j$, the dipole-dipole interaction
Hamiltonian reads \citep{SWM10} 
\begin{equation}
V_{\text{dd}}^{\left(i,j\right)}\left(\vec{r}_{ij}\right)=\vec{\text{d}}^{\left(i\right)}\vec{\text{d}}^{\left(j\right)}/r_{ij}^{3}-3\left(\vec{\text{d}}^{\left(i\right)}\vec{r}_{ij}\right)\left(\vec{\text{d}}^{\left(j\right)}\vec{r}_{ij}\right)/r_{ij}^{5},\label{eq:Vdd}
\end{equation}
where $\vec{\text{d}}^{(i)}$ is the dipole operator of $i$-th atom
and $\vec{r}_{ij}$ is the relative distance between atoms. In second-order
perturbation theory in $\hat{V}_{\text{dd}}^{\left(i,j\right)}$ we
obtain the effective van der Waals interaction \citep{VGZ15,GDN15}
(we assume the absence of Förster resonances\citep{Ravets:2014aa})

\begin{equation}
H_{\text{vdW}}^{\left(i,j\right)}\equiv P\sum_{\beta,\chi}\frac{V_{dd}^{\left(i,j\right)}Q_{\beta\chi}V_{dd}^{\left(i,j\right)}}{\delta_{\beta\chi}}P,\label{eq:vdW}
\end{equation}
Here $\hat{P}$ is the projector onto the states of interest ($nP_{1/2},\,nP_{1/2}$
or $nP_{1/2},\,n^{\prime}S_{1/2}$ ), and $Q_{\beta\chi}\equiv\left|\beta,\chi\right\rangle \left\langle \beta,\chi\right|$
is the projector on manifolds that are populated only as virtual intermediate
states, with $\delta_{\beta\chi}$ energy differences. We note that
due to the perturbative nature of~\eqref{eq:vdW}, this expression
is valid only beyond a certain critical radius $r>r_{c}$ (see below). 

\subsection{Interaction between simulator atoms\label{p-p_interaction}}

\begin{figure}
\begin{centering}
\includegraphics[scale=0.38]{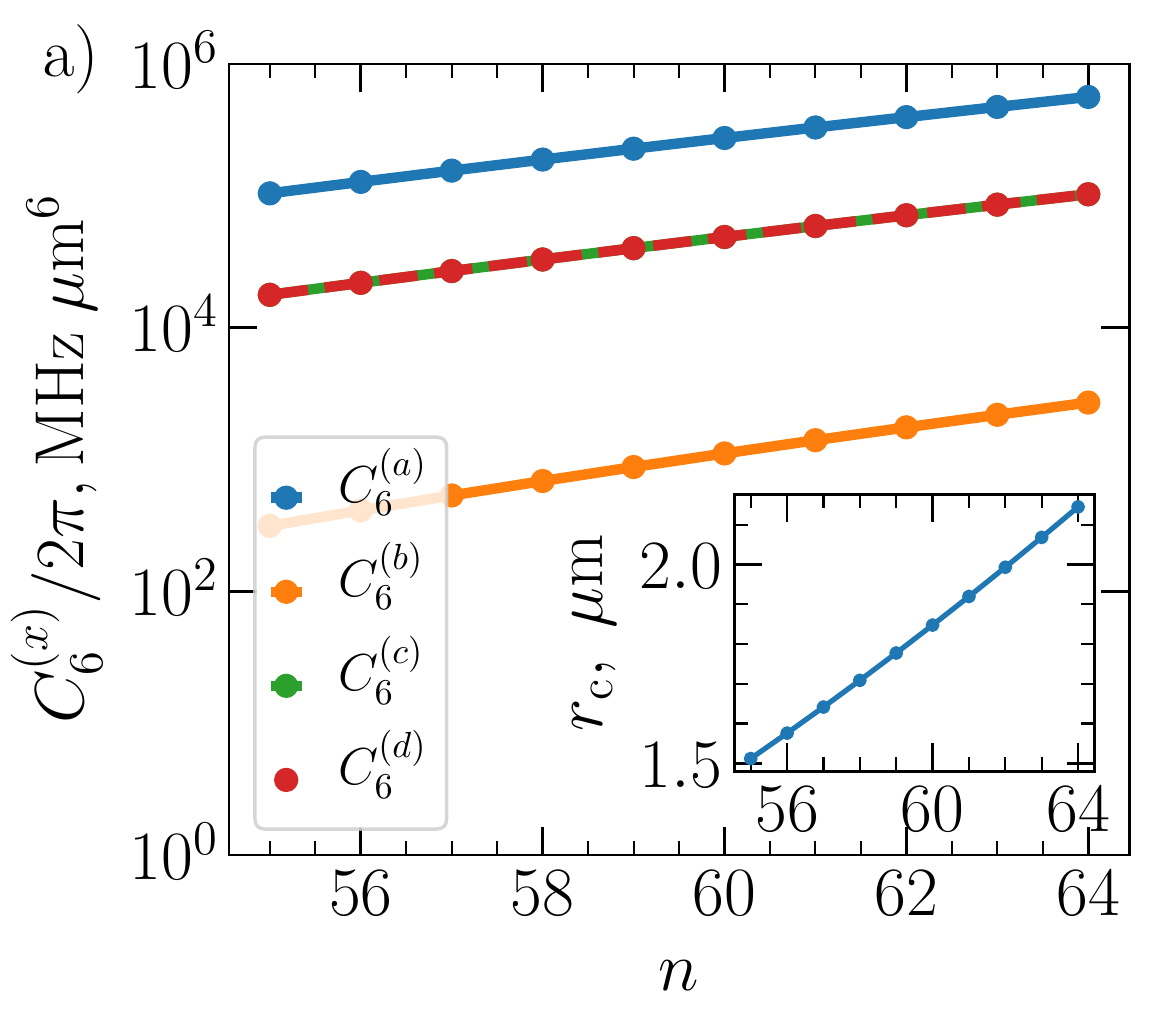}\includegraphics[scale=0.38]{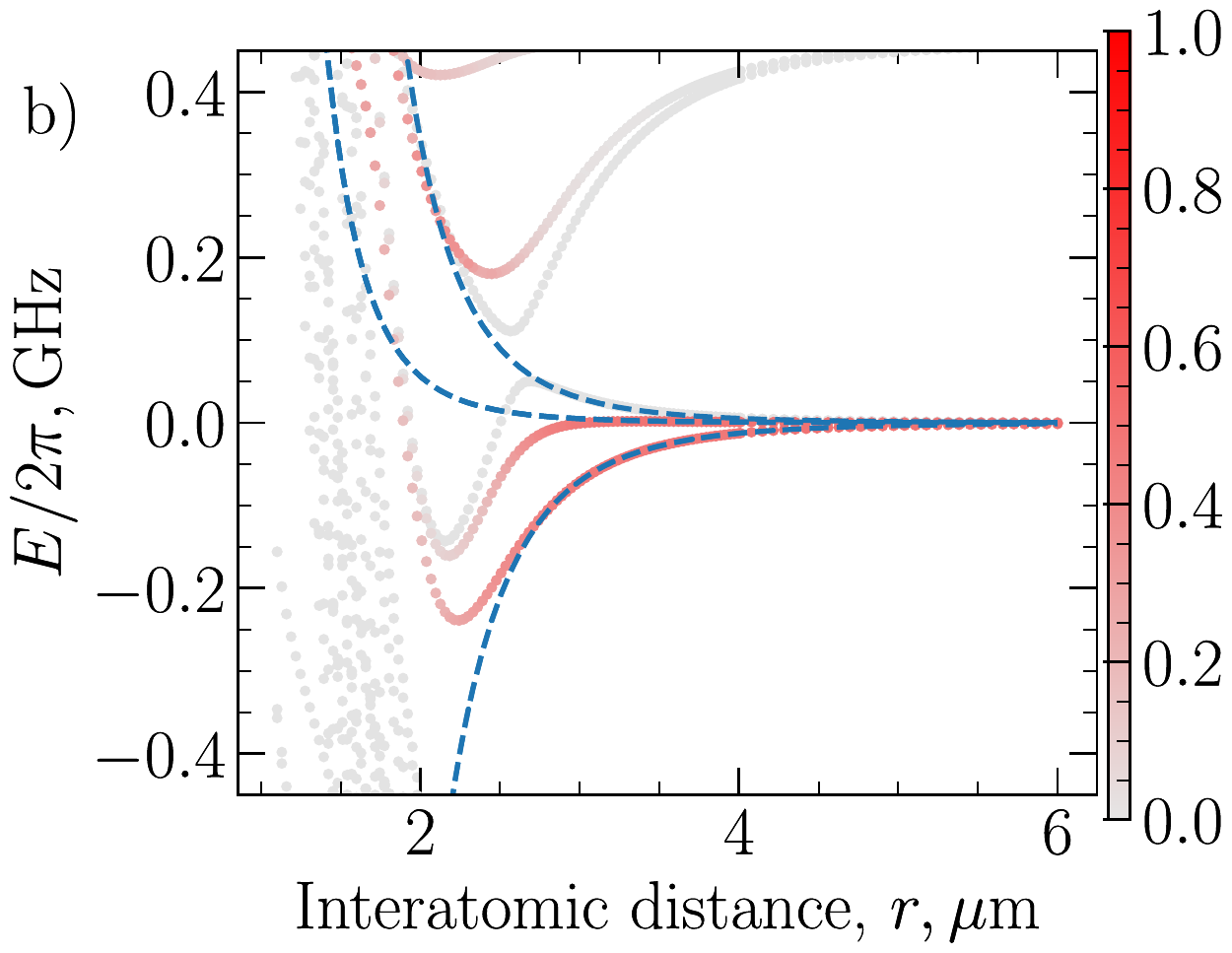}
\par\end{centering}
\caption{Van der Waals interaction of two $nP_{1/2}$ Rydberg states. (a) $C_{6}^{\left(x\right)}$
coefficient as function of the principal quantum number $n$. Inset
shows the critical radius $r_{\text{c}}$. (b) Exact diagonalization
of dipole-dipole interaction Eq.~\eqref{eq:Vdd}. Color indicates
the projection of exact eigenstates onto the state $\left|r_{+}\right\rangle _{i}\left|r_{+}\right\rangle _{j}$.
Blue dashed line stands for the eigenvalues of the perturbative Hamiltonian
Eq.~\eqref{eq:Vdw_1}.}

\label{Fig3}
\end{figure}

We first consider the van der Waals interaction between excited Rydberg
states for the fine structure states $\left|r_{\alpha=\pm}\right\rangle =\left|nP_{1/2},m_{J}=\pm1/2\right\rangle $
of simulator atoms. The relevant projector reads

\[
P=\sum_{\alpha=\pm}\left|r_{\alpha}\right\rangle _{i}\left\langle r_{\alpha}\right|\otimes\sum_{\alpha=\pm}\left|r_{\alpha}\right\rangle _{j}\left\langle r_{\alpha}\right|.
\]
\begin{table}[H]
\begin{centering}
\begin{tabular}{cccc}
\hline 
a) & $nP_{1/2}+nP_{1/2}$ & $\leftrightarrow$ & $n_{\beta}S_{1/2}+n_{\chi}S_{1/2}$\tabularnewline
\hline 
b) & $nP_{1/2}+nP_{1/2}$ & $\leftrightarrow$ & $n_{\beta}D_{3/2}+n_{\chi}D_{3/2}$\tabularnewline
\hline 
c) & $nP_{1/2}+nP_{1/2}$ & $\leftrightarrow$ & $n_{\beta}D_{3/2}+n_{\chi}S_{1/2}$\tabularnewline
\hline 
d) & $nP_{1/2}+nP_{1/2}$ & $\leftrightarrow$ & $n_{\beta}S_{1/2}+n_{\chi}D_{3/2}$\tabularnewline
\hline 
\end{tabular}
\par\end{centering}
\caption{Channels of the dipole-dipole interaction of $nP_{1/2}$ Rydberg states.}

\label{Channels_1}
\end{table}
The possible interaction channels for the atoms in $P_{1/2}$ states
are listed in Table~\ref{Channels_1}, which provides us with the
intermediate states $\left|\beta,\chi\right\rangle $ of Eq.~\eqref{eq:vdW}.
This allows us to write the matrix of van der Waals interaction Hamiltonian
$H_{\text{vdW}}^{\left(i,j\right)}$ in the form~\eqref{eq:V_vdw-2}
\citep{BYB19} with $C_{6}=2\left[C_{6}^{\left(a\right)}+4C_{6}^{\left(b\right)}+2\left(C_{6}^{\left(c\right)}+C_{6}^{\left(d\right)}\right)\right]/27$
and $\widetilde{C}_{6}=C_{6}^{\left(a\right)}+C_{6}^{\left(b\right)}-C_{6}^{\left(c\right)}-C_{6}^{\left(d\right)}$
combinations of coefficients $C_{6}^{\left(a,b,c,d\right)}$ attributed
to different scattering channels. (Explicit expressions involving
Clebsch-Gordan and dipole matrix elements can be found in \citep{VGZ15,GDN15}),
and corresponding plots for Rb atoms are shown in Fig.~\ref{Fig3}(a).
The $4\times4$ matrix $\mathbb{D}_{0}\left(\theta,\phi\right)$ referred
to in Eq.~\eqref{eq:V_vdw-2} reads

\begin{align*}
 & \mathbb{D}_{0}\left(\theta,\phi\right)=\\
 & \left(\begin{array}{cccc}
\frac{3\cos\left(2\theta\right)-1}{81} & \frac{4e^{-i\phi}\sin\left(2\theta\right)}{27} & \frac{4e^{-i\phi}\sin\left(2\theta\right)}{27} & \frac{2e^{-2i\phi}\sin^{2}\left(\theta\right)}{27}\\
\frac{e^{i\phi}\sin\left(2\theta\right)}{27} & \frac{1-3\cos\left(2\theta\right)}{81} & \frac{-5-3\cos\left(2\theta\right)}{81} & \frac{-4e^{-i\phi}\sin\left(2\theta\right)}{27}\\
\frac{e^{i\phi}\sin\left(2\theta\right)}{27} & \frac{-5-3\cos\left(2\theta\right)}{81} & \frac{1-3\cos\left(2\theta\right)}{81} & \frac{-4e^{-i\phi}\sin\left(2\theta\right)}{27}\\
\frac{2e^{2i\phi}\sin^{2}\left(\theta\right)}{27} & \frac{-4e^{i\phi}\sin\left(2\theta\right)}{27} & \frac{-4e^{i\phi}\sin\left(2\theta\right)}{27} & \frac{3\cos\left(2\theta\right)-1}{81}
\end{array}\right),
\end{align*}
and following the geometry of our setup in Fig.~\ref{fig:1}(b) we
set $\theta=\pi/2$. Thus we obtain for the interaction Hamiltonian
the structure

\begin{equation}
V_{\text{vdW}}^{\left(i,j\right)}=\left(\begin{array}{cccc}
W_{++}^{\left(i,j\right)} & 0 & 0 & V_{++}^{\left(i,j\right)}\\
0 & W_{+-}^{\left(i,j\right)} & V_{-+}^{\left(i,j\right)} & 0\\
0 & V_{+-}^{\left(i,j\right)} & W_{-+}^{\left(i,j\right)} & 0\\
V_{--}^{\left(i,j\right)} & 0 & 0 & W_{--}^{\left(i,j\right)}
\end{array}\right),\label{eq:Vdw_1}
\end{equation}
where
\[
W_{++}^{\left(i,j\right)}=W_{--}^{\left(i,j\right)}=\frac{1}{r_{ij}^{6}}[C_{6}-\frac{4}{81}\widetilde{C}_{6}],
\]
\[
W_{+-}^{\left(i,j\right)}=W_{-+}^{\left(i,j\right)}=\frac{1}{r_{ij}^{6}}[C_{6}+\frac{4}{81}\widetilde{C}_{6}],
\]
\[
V_{+-}^{\left(i,j\right)}=V_{-+}^{\left(i,j\right)}=-\frac{2}{81}\frac{\widetilde{C}_{6}}{r_{ij}^{6}},
\]
and
\begin{equation}
V_{++}^{\left(i,j\right)}=V_{--}^{\left(i,j\right)*}=-\frac{2}{27}\frac{\widetilde{C}_{6}}{r_{ij}^{6}}\exp(-2i\phi).\label{eq:Vpp}
\end{equation}
The specific spin models which can be engineered via Rydberg dressing
\citep{GDN15,Bijnen2015}, i.e. by admixing the van der Waals interactions~\eqref{eq:Vdw_1}
to the ground states by off-resonant laser light, is determined by
the structure of the matrix elements~\eqref{eq:Vpp}. 

\subsection{Interactions between simulator and control atoms\label{p-s_interaction}}

\begin{figure}
\begin{centering}
\includegraphics[width=1\columnwidth]{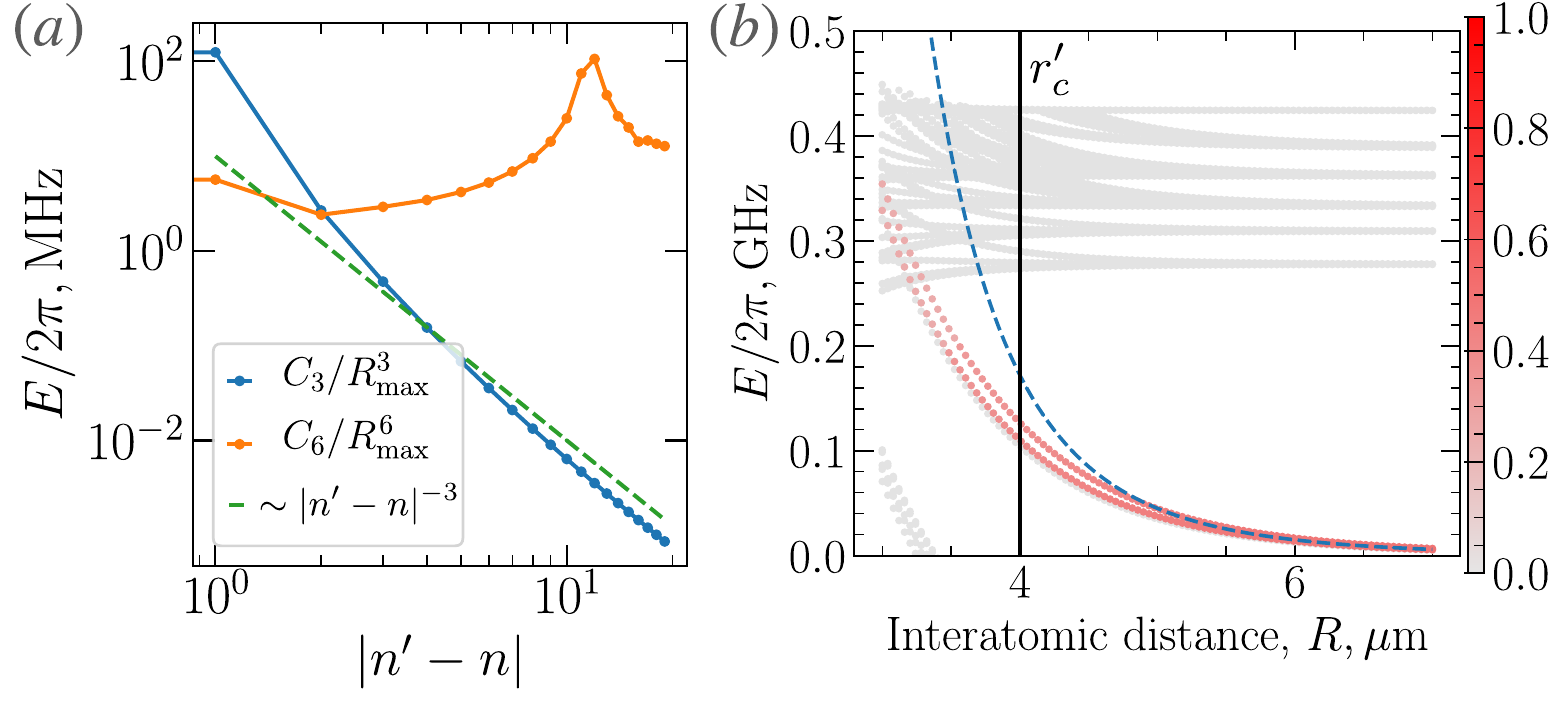}
\par\end{centering}
\caption{Interaction energy between the control and simulator atoms in the
Rydberg states. (a) Comparison of the strength of the direct dipole-dipole
(blue) and the van der Waals (orange) interactions at $R_{{\rm max}}=4.6\mu\text{m}$
for different values of $n'$ and $n=60$. (b) Exact diagonalization
of the dipole-dipole interaction Eq.~\eqref{eq:Vdd}. Color indicates
the projection of exact eigenstates onto the state $\left|1\right\rangle _{c}\left|r_{+}\right\rangle _{j}$.
Blue dashed line stands for the eigenvalues of the perturbative Hamiltonian
Eq.~\eqref{eq:vdw_control-1}.}

\label{Fig_validity}
\end{figure}

Here we consider the van der Waals interaction between the simulator
and control atoms in the Rydberg states. We restrict the states of
the control atom to the Zeeman manifold which includes the logical
qubit state $\left|1\right\rangle _{c}$, i.e.~$\left|1\right\rangle _{c}=\left|n^{\prime}S_{1/2},m_{j}=1/2\right\rangle $,
and $\left|1^{\prime}\right\rangle _{c}\equiv\left|n^{\prime}S_{1/2},m_{j}=-1/2\right\rangle $.
As we discuss below, the unwanted coupling between $\left|1\right\rangle _{c}$
and $\left|1^{\prime}\right\rangle _{c}$ can be minimized by proper
choice of the principal quantum number $n^{\prime}$ \citep{BYB19}.
For the states of simulator atoms we again consider $\left|r_{\alpha=\pm}\right\rangle =\left|nP_{1/2},m_{J}=\pm1/2\right\rangle $.
As already mentioned in the main text, we assume $\left|n-n^{\prime}\right|\gg1$
in order to avoid direct dipolar exchange interaction between the
control and simulator atoms. Comparison of the typical values of the
van der Waals and dipolar interaction strengths are shown in Fig.~\ref{Fig_validity}(a)
for distances $R_{\text{max}}\approx4.6\mu\text{m}$.

\begin{table}[H]
\begin{centering}
\begin{tabular}{cccc}
\hline 
a) & $nP_{1/2}+n^{\prime}S_{1/2}$ & $\leftrightarrow$ & $n_{\beta}S_{1/2}+n_{\chi}P_{1/2}$\tabularnewline
\hline 
b) & $nP_{1/2}+n^{\prime}S_{1/2}$ & $\leftrightarrow$ & $n_{\beta}D_{3/2}+n_{\chi}P_{3/2}$\tabularnewline
\hline 
c) & $nP_{1/2}+n^{\prime}S_{1/2}$ & $\leftrightarrow$ & $n_{\beta}D_{3/2}+n_{\chi}P_{1/2}$\tabularnewline
\hline 
d) & $nP_{1/2}+n^{\prime}S_{1/2}$ & $\leftrightarrow$ & $n_{\beta}S_{1/2}+n_{\chi}D_{3/2}$\tabularnewline
\hline 
\end{tabular}
\par\end{centering}
\caption{Channels of the dipole-dipole interaction of $nP_{1/2},n^{\prime}S_{1/2}$
Rydberg states.}

\label{Channels_1-1}
\end{table}

The relevant scattering channels are listed in Table~\ref{Channels_1-1}.
Similar to~\eqref{p-p_interaction} above, we obtain for $H_{\text{vdW}}^{\left(i,c\right)}$
in the basis ($\left|r_{+}\right\rangle \left|1\right\rangle _{\text{c}}$,
$\left|r_{+}\right\rangle \left|1^{\prime}\right\rangle _{\text{c}}$,
$\left|r_{-}\right\rangle \left|1\right\rangle _{\text{c}}$, $\left|r_{-}\right\rangle \left|1^{\prime}\right\rangle _{\text{c}}$)
an expression analogous to Eq.~\eqref{eq:V_vdw-2},

\begin{align}
V_{\text{vdW}}^{\left(i,c\right)}= & \frac{1}{r_{i,c}^{6}}\left[C_{6}^{\prime}\mathbb{I}_{4}-\widetilde{C}_{6}^{\prime}\mathbb{D}_{0}\left(\theta,\phi\right)\right],\label{eq:V_vdw-1}
\end{align}
where coefficients $C_{6}^{\prime}$ and $\widetilde{C}_{6}^{\prime}$
are defined as above. Numerical results for $C_{6}^{\prime}$, and
the relative strength of the second (anisotropic) term in Eq.~\eqref{eq:V_vdw-1}
are shown in Figs.~\ref{Fig4}(a,b), respectively, for different
principle quantum numbers $n$ and $n'=n+\Delta n$. As can be seen
from Fig.~\ref{Fig4}(b), the second term in $V_{\text{vdW}}^{\left(i,c\right)}$
can be made much smaller than the first one by choosing $n$ and $\Delta n$
properly. This makes the interaction essentially isotropic. 

We note, however, that, even when the condition $\left|\widetilde{C}_{6}^{\prime}\right|\ll\left|C_{6}^{\prime}\right|$
is satisfied, the second term in Eq.~\eqref{eq:V_vdw-1} may cause
unwanted transitions between the Rydberg states of the control atom,
$\left|1\right\rangle _{\text{c}}\leftrightarrow\left|1^{\prime}\right\rangle _{\text{c}}$,
which can bring the control atom out of the Hilbert space of interest.
To suppress such transitions, we make them strongly off-resonant by
imposing e.g.~an external magnetic field. Thus we assume that the
Hilbert space of the control atom can be represented by the two qubit
states $\left|0\right\rangle _{\text{c}}$ and $\left|1\right\rangle _{\text{c}}$,
and we can describe the interaction between the simulator atom and
the c-qubit by the `blockade' Hamiltonian~\eqref{eq:vdw_control-1}.
The comparison with the exact diagonalization is shown in Fig.~\ref{Fig_validity}(b).
\begin{center}
\begin{figure}
\begin{centering}
\includegraphics[scale=0.5]{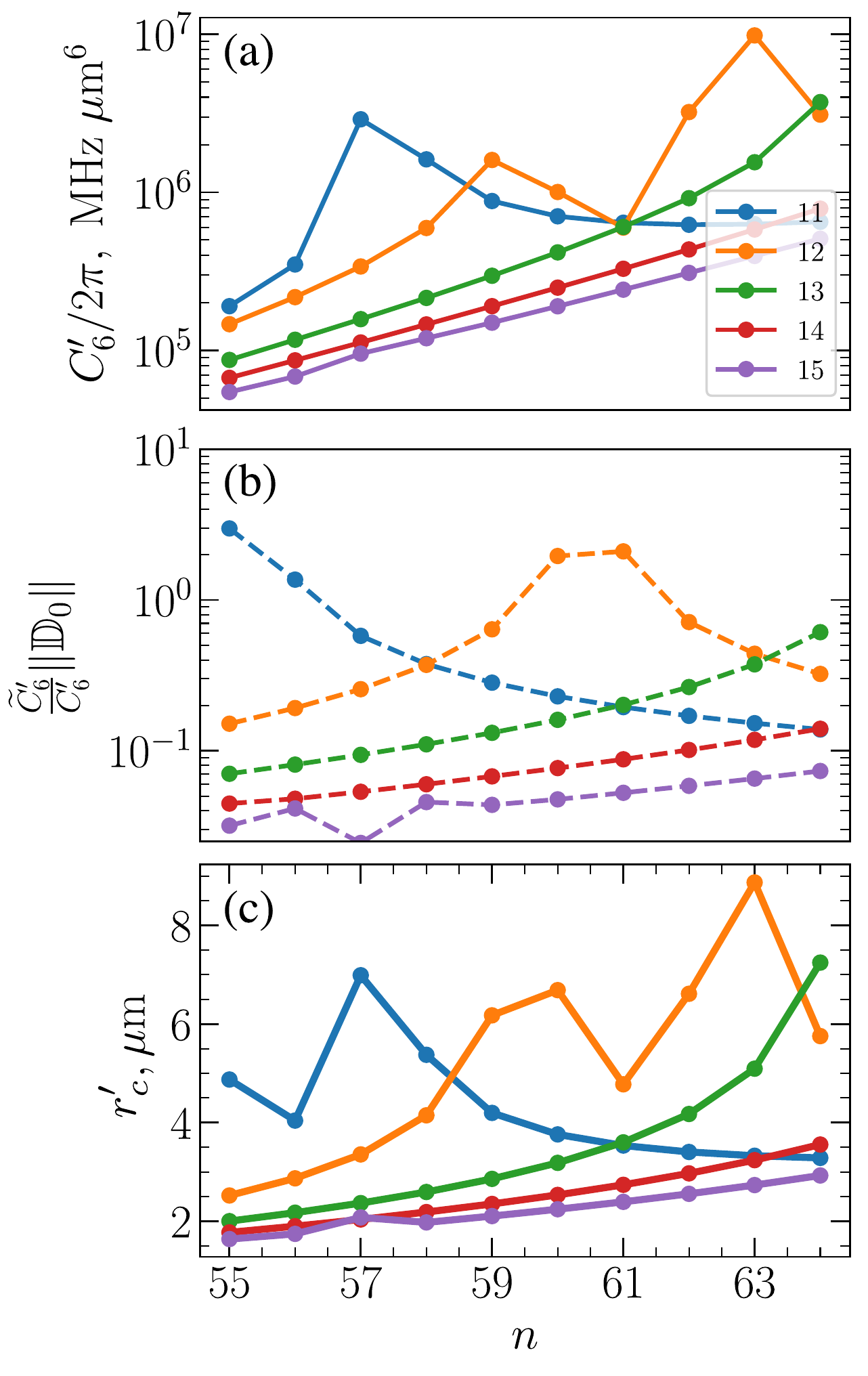}
\par\end{centering}
\caption{Van der Waals interaction between the Rydberg states of the simulator
and control atoms with the principle quantum numbers $n$ and $n'=n+\Delta n$,
respectively ($\Delta n$ is shown in different colors as indicated
on the inset). (a) $C_{6}^{\prime}$ coefficient as a function of
$n$ for different values of $\Delta n$. (b) Relative strength of
two contribution to $V_{\text{vdW}}^{\left(i,c\right)}$ with $\left\Vert \mathbb{D}_{0}\right\Vert =2\sqrt{2}/27$
being the Frobenius norm of the matrix $\mathbb{D}_{0}$. (c) Critical
distance $r_{c}'$ for the van der Waals interaction between simulator
and control atoms, see text.}

\label{Fig4}
\end{figure}
\par\end{center}

\subsection{Validity of the van der Waals Hamiltonian\label{subsec:Validity-of-the}}

Before proceeding, we emphasize that the validity of the van der Waals
interaction Hamiltonian Eq.~\eqref{eq:vdW} requires the conditions
\citep{SWM10,B13} to be satisfied:

\begin{align*}
\max_{\beta\chi}\left\{ _{i}\left\langle r_{\pm}\right|{}_{j}\left\langle r_{\pm}\right|V_{dd}^{\left(i,j\right)}\delta_{\beta\chi}^{-2}Q_{\beta\chi}V_{dd}^{\left(i,j\right)}\left|r_{\pm}\right\rangle _{i}\left|r_{\pm}\right\rangle _{j}\right\}  & \ll1,\\
\max_{\beta\chi}\left\{ _{\text{c}}\left\langle 1\right|{}_{i}\left\langle r_{\pm}\right|V_{dd}^{\left(i,j\right)}\delta_{\beta\chi}^{-2}Q_{\beta\chi}V_{dd}^{\left(i,j\right)}\left|r_{\pm}\right\rangle _{i}\left|1\right\rangle _{\text{c}}\right\}  & \ll1
\end{align*}
which sets the lower bound on the interatomic distance, $r_{ij}>r_{c}$,
for $i,j=1,\ldots,L$, and $r_{i,c}>r_{c}'$, where the critical distances
$r_{c}$ and $r_{c}'$ are defined as the distances when the left-hand-site
in the above condition equals unity. The calculated dependencies of
$r_{c}$ and $r_{c}'$ on the main quantum numbers are shown in the
inset of Fig.~\ref{Fig3}(a) and in Fig.~\ref{Fig4}(c), respectively.
For the simulator atoms, however, direct diagonalization of $V_{\text{dd}}^{\left(i,j\right)}$,
Eq.~\eqref{eq:Vdd}, shows the presence of an avoided crossing at
$r_{ij}\approx2.4\mu{\rm m},$ see Fig.~\ref{Fig3}(b). Therefore,
to stay away from this nonperturbative situation, we take $r_{c}=2.4\mu{\rm m}$
and (using $n=60$ and $n^{\prime}=71$ as principal quantum numbers)\textbf{
}for the shortest interatomic distances in our setup.

\section{Derivation of the effective Hamiltonian\label{sec:Appendix.Adiabatic-elimination-of}}

In this Appendix we derive the effective Hamiltonian of the simulator-qubit
interaction which we show to be given by Eq.~\eqref{eq:H_system-qubit-1}.
We perform an adiabatic elimination of excited Rydberg states of simulator
atoms $\left|r_{\pm}\right\rangle $ following \citep{GDN15}. We
first transform Hamiltonian $H_{\text{s}}$ {[}see Eqs.~\eqref{eq:H_s}-\eqref{eq:H_sc}{]}
into rotating frame:

\begin{align}
H_{s} & =\sum_{i=1}^{L}H_{0}^{(i)}+\sum_{i,j=1}^{L}H_{\text{vdW}}^{\left(i,j\right)},\label{eq:H_rf-1}
\end{align}
where

\begin{align}
H_{0}^{(i)}= & \sum_{\alpha=\pm}(\Omega_{\alpha}\left|g_{\alpha}\right\rangle \left\langle g_{\alpha}\right|+\text{H.c.})\nonumber \\
- & \Delta_{B}(\left|r_{-}\right\rangle _{i}\left\langle r_{-}\right|+\left|g_{-}\right\rangle _{i}\left\langle g_{-}\right|)\nonumber \\
- & \sum_{\alpha=\pm}\Delta_{\alpha}\left|r_{\alpha}\right\rangle _{i}\left\langle r_{\alpha}\right|,\label{eq:18-1}
\end{align}
and $\Delta_{B}=\omega_{+}-\omega_{-}+E_{g_{+}}-E_{g_{-}}$, $\Delta_{\pm}=E_{r\pm}-E_{g\pm}-\omega_{\pm}$.
The explicit expression of van der Waals interaction Hamiltonian according
to Eq.~\eqref{eq:Vdw_1} is 

\begin{align*}
H_{\text{vdW}}^{\left(i,j\right)}= & \sum_{\alpha,\alpha^{\prime}}W_{\alpha\alpha^{\prime}}^{\left(i,j\right)}\left|r_{\alpha}\right\rangle _{i}\left\langle r_{\alpha}\right|\otimes\left|r_{\alpha^{\prime}}\right\rangle _{j}\left\langle r_{\alpha^{\prime}}\right|\\
+ & V_{+-}^{\left(i,j\right)}\left|r_{+}\right\rangle _{i}\left\langle r_{-}\right|\otimes\left|r_{-}\right\rangle _{j}\left\langle r_{+}\right|+\text{H.c.}\\
+ & V_{--}^{\left(i,j\right)}\left|r_{+}\right\rangle _{i}\left\langle r_{-}\right|\otimes\left|r_{+}\right\rangle _{j}\left\langle r_{-}\right|+\text{H.c.,}
\end{align*}
We now define the dressed ground $\left|\tilde{g}_{\alpha}\right\rangle _{i}$
and Rydberg $\left|\tilde{r}_{\alpha}\right\rangle _{i}$ states as
the eigenstates of the Hamiltonian $H_{0}^{\left(i\right)}$,
\begin{equation}
\begin{aligned}\left|\tilde{g}_{\alpha}\right\rangle _{i}= & \frac{\sqrt{\lambda_{-}}\ket{g_{\alpha}}_{i}-\sqrt{\lambda_{+}}\ket{r_{\alpha}}_{i}}{\sqrt{\lambda_{+}+\lambda_{-}}},\\
\left|\tilde{r}_{\alpha}\right\rangle _{i}= & \frac{\sqrt{\lambda_{+}}\ket{g_{\alpha}}_{i}+\sqrt{\lambda_{-}}\ket{r_{\alpha}}_{i}}{\sqrt{\lambda_{+}+\lambda_{-}}},
\end{aligned}
\label{eq: dressed states}
\end{equation}
where $\lambda_{\pm}=$$[(\Delta_{\alpha}/2)^{2}+\Omega_{\alpha}^{2}]^{1/2}\pm\Delta_{\alpha}/2$
(we assume $\Delta_{\alpha}>0$ and real Rabi frequencies $\Omega_{\alpha}$).

We now define the projectors onto the subspace of the dressed ground
states $\left|\tilde{g}_{\alpha}\right\rangle _{i}$

\begin{align}
P_{g} & =\bigotimes_{i=1}^{L}\left(\sum_{\alpha=\pm}\left|\tilde{g}_{\alpha}\right\rangle _{i}\left\langle \tilde{g}_{\alpha}\right|\right)\label{eq:P_g}
\end{align}
and $Q=\bigotimes_{i=1}^{L}\mathbb{I}_{i}-P_{g}.$ Equivalently, we
can write $P_{g}$ as a sum of projectors onto the eigenstates of
the Hamiltonian $\sum_{i}H_{0}^{\left(i\right)}$ with the eigenenergy
$E_{l}^{(0)}$, $P_{g}=\sum_{l}P_{g}^{\left(l\right)}$, where $P_{g}^{\left(l\right)}$
is the projector onto the subspace formed by the eigenstate with the
eigenenergy $E_{l}^{\left(0\right)}$. Due to the particular form
of the blockade interaction Eq.~\eqref{eq:V_vdw-1} we can perform
the adiabatic elimination of excited states simultaneously for the
two states of the qubit. We now derive the effective Hamiltonian of
the simulator atoms using projectors, and write an effective Hamiltonian
in the $P_{g}$ subspace up to two lowest orders in $V$ as \citep{RS12}\begin{widetext}

\begin{equation}
H_{\text{eff}}=P_{g}H_{\text{s}}P_{g}+\frac{1}{2}\left\{ P_{g}H_{\text{s}}Q\sum_{l}\left(E_{l}^{(0)}-QH_{\text{s}}Q\right)^{-1}QH_{\text{s}}P_{g}^{\left(l\right)}+\text{H.c.}\right\} .\label{eq:PHP}
\end{equation}
\end{widetext} After evaluating this expression analytically up to
the 4-th order in $\xi_{\alpha}\equiv\Omega_{\alpha}/\Delta_{\alpha}$,
we obtain \citep{GDN15} the effective Hamiltonian in general XYZ
form:
\begin{align}
H_{\text{eff}}(\Delta_{\alpha})= & \sum_{i<j=1}^{L}\sum_{\eta=x,y,z}J_{ij}^{\left(\eta\right)}\left(\Delta_{\alpha}\right)\sigma_{i}^{\eta}\sigma_{j}^{\eta}\nonumber \\
+ & \sum_{i=1}^{L}\left[h^{z}\left(\Delta_{\alpha}\right)+\frac{1}{2}\Delta_{B}\right]\sigma_{i}^{z},\label{eq:JxJyJz}
\end{align}

We now show that the model can be reduced to the Heisenberg XXZ model
assumed in the main text, and provide the corresponding interaction
coefficients $J_{ij}^{\left(\eta\right)},h^{z}$. The expressions
for the Hamiltonians $H_{\text{spin}}$ and $H_{\text{spin}}^{\prime}$
in Eq.~\eqref{eq:H_system-qubit-1} are

\begin{align}
H_{\text{spin}} & =H_{\text{eff}}\left(\Delta_{\alpha}\right),\label{eq:Heff_spin}\\
H_{\text{spin}}^{\prime} & =H_{\text{eff}}\left(\Delta_{\alpha}-C_{6}^{\prime}/R^{6}\right)\label{eq:Heffp_spin}
\end{align}
As a result of the adiabatic elimination procedure the qubit logical
states $\left|0\right\rangle _{\text{c}}$ and $\left|1\right\rangle _{\text{c}}$
respectively acquire additional spin-independent shifts $\sum\beta_{ij}$
and $\sum\beta_{ij}^{\prime}$ (see below). Thus in the rotating frame
of the control laser frequency, the Hamiltonian takes on the form
Eq.~\eqref{eq:H_system-qubit-1} with an effective detuning $\delta=\sum_{ij}\left(\beta_{ij}^{\prime}-\beta_{ij}^{\prime}\right)+E_{\text{cr}}-E_{\text{cg}}$.

We now discuss the effective Hamiltonians~\eqref{eq:Heff_spin} and
\eqref{eq:Heffp_spin}. In the case of a perfect blockade, $C_{6}^{\prime}/\left(\Delta_{\alpha}R^{6}\right)\rightarrow\infty$,
one has $J_{ij}^{\left(\eta\right)}\left(\Delta_{\alpha}-C_{6}^{\prime}/R^{6}\right)\rightarrow0$
and $h_{i}^{z}\left(\Delta_{\alpha}-C_{6}^{\prime}/R^{6}\right)\rightarrow0$
and, therefore, $H_{\text{spin}}^{\prime}$ reduces to a simple form
$H_{\text{spin}}^{\prime}=(1/2)\Delta_{B}\sum_{i=1}^{L}\sigma_{i}^{z}$.
With this we are able to identify two regimes when the Hamiltonian
Eq.~\eqref{eq:H_system-qubit-1} gives rise to the desired QND Hamiltonian,
i.e.~$[H_{{\rm spin}},H_{{\rm spin}}^{\prime}]=0$. The first one
corresponds to $\Delta_{B}=0$. In this manuscript we focus on the
second regime of a ``strong magnetic field'' when $\Delta_{B}\rightarrow\infty$
(more precisely, $\left|\Delta_{B}\right|\gg|J_{ij}^{\left(\eta\right)}|,|h_{i}^{z}|$)
when the spin model effectively reduces to the Heisenberg XXZ chain.
The explicit expressions for the couplings are then given by\begin{widetext}

\begin{align}
h^{z}= & 0,\label{eq:hz}\\
J_{ij}^{\left(x,y\right)}= & -2\Delta^{2}\xi^{4}\frac{V_{+-}^{\left(i,j\right)}}{\left(V_{+-}^{\left(i,j\right)}-W_{+-}^{\left(i,j\right)}-2\Delta\right)\left(2\Delta+V_{+-}^{\left(i,j\right)}+W_{+-}^{\left(i,j\right)}\right)},\label{eq:Jxy}\\
J_{ij}^{\left(z\right)}= & -2\Delta^{2}\xi^{4}\frac{V_{+-}^{2}-\left(2\Delta+W_{+-}^{\left(i,j\right)}\right)\left(W_{+-}^{\left(i,j\right)}-W_{++}^{\left(i,j\right)}\right)}{\left(2\Delta+W_{+-}^{\left(i,j\right)}\right)\left(V_{+-}^{\left(i,j\right)}-W_{+-}^{\left(i,j\right)}-2\Delta\right)\left(2\Delta+V_{+-}^{\left(i,j\right)}+W_{+-}^{\left(i,j\right)}\right)},\label{eq:Jz}\\
\beta_{ij}= & -2\Delta\xi^{2}\nonumber \\
+ & 2\Delta\xi^{4}\frac{V_{+-}^{2}\left(3\Delta+2W_{++}^{\left(i,j\right)}\right)-\left(2\Delta+W_{+-}^{\left(i,j\right)}\right)\left[4\Delta^{2}+3\Delta\left(W_{+-}^{\left(i,j\right)}+W_{++}^{\left(i,j\right)}\right)+2W_{+-}^{\left(i,j\right)}W_{++}^{\left(i,j\right)}\right]}{\left(2\Delta+W_{++}^{\left(i,j\right)}\right)\left(V_{+-}^{\left(i,j\right)}-2\Delta-W_{+-}^{\left(i,j\right)}\right)\left(2\Delta+V_{+-}^{\left(i,j\right)}+W_{+-}^{\left(i,j\right)}\right)}\label{eq:beta_ij}
\end{align}
\end{widetext} where $V$ and $W$ refer to the van der Waals interaction
derived in~\eqref{p-p_interaction}, and we assumed equal Rabi frequencies
and detunings, such that $\xi_{\alpha}=\xi$. If the energy difference
between the subspaces is much larger than the coupling $|J_{ij}^{\left(\eta\right)}|\ll\left|\Delta_{B}\right|$,
this term can be neglected, such that the resulting low-energy dynamics
(with typical energies $\sim J$) takes place in a the sector with
a fixed $S_{z}=\sum_{i}\sigma_{i}^{z}$.

To conclude, we note that in Eq.~\eqref{eq:JxJyJz} one can also
generate a\emph{ position-dependent} magnetic field, $h^{z}\to h_{i}^{z}=h^{z}+\delta h_{i}^{z}$.
This can be achieved by using position dependent Rabi frequencies,
$\Omega_{\alpha}^{\left(i\right)}=\Omega_{\alpha}+\delta\Omega_{\alpha}^{\left(i\right)}$,
which results in $\delta h_{i}^{z}\approx\sum_{\alpha}\alpha\xi_{\alpha}\delta\Omega_{\alpha}^{\left(i\right)}$.
After choosing $\delta\Omega_{\alpha}^{\left(i\right)}\sim(J/\xi)w_{i}$,
where $w_{i}\in[-1,1]$ is a uniformly distributed random number,
we obtain $\delta h_{i}^{z}\sim J$.

\section{Implementation of complex flip-flop phases\label{subsec:Complex J}}

We now provide a way to engineer \emph{complex} flip-flop coefficients
in our spin models. As a starting point, let us consider the dressing
lasers with the Laguerre-Gaussian spatial mode profile corresponding
to the angular momentum $l_{\pm}$ (note that it is sufficient to
have only one nonzero $l_{\alpha}$). In this case, the dressing term
in the Hamiltonian Eq.~\eqref{eq:H_s} becomes

\begin{align}
\sum_{k=1}^{L}\sum_{\alpha=\pm}\left(\Omega_{\alpha}^{\mathrm{LG}}e^{i\phi_{k,\alpha}}\left|g_{\alpha}\right\rangle _{k}\left\langle r_{\alpha}\right|e^{i\omega_{\alpha}t}+\text{H.c.}\right),
\end{align}
where $\phi_{k,\pm}\equiv2\pi kl_{\pm}/L$. After performing the adiabatic
elimination procedure as in Sec.~\ref{sec:Appendix.Adiabatic-elimination-of}
of this Appendix, we obtain the effective Hamiltonian of the form
following form (we omit here the $\Delta_{B}$-term)

\begin{align}
 & H_{\text{eff}}\left(\tilde{\Delta}_{\alpha}\right)\nonumber \\
 & =\sum_{i<j=1}^{L}\left\{ 2J_{kj}^{\left(x,y\right)}\left(e^{i\sum_{\alpha}\alpha\left(\phi_{k,\alpha}-\phi_{j,\alpha}\right)}\sigma_{k}^{+}\sigma_{j}^{-}+\text{H.c.}\right)\right\} \nonumber \\
 & +\sum_{i<j=1}^{L}\left\{ J_{kj}^{\left(z\right)}\sigma_{k}^{z}\sigma_{j}^{z}\right\} +\sum_{i=1}^{L}h_{i}^{z}\sigma_{i}^{z},\label{eq:Heff0-1}
\end{align}
with the flip-flop amplitudes acquiring nonzero phases $\phi_{kj}=i\sum\alpha\left(\phi_{k,\alpha}-\phi_{j,\alpha}\right)$.
We note, however, that the sum of the flip-flop phases in Eq.~\eqref{eq:Heff0-1}
across the system is always equal to an integer number of $2\pi$,
having, therefore, no effect on spectral statistics. However, physically
relevant phase of the flip-flop amplitudes can be generated using
stroboscopic engineering. Let us consider the case when, during the
stroboscopic period $T=2\pi/\omega_{s}$, we use the plane-wave dressing
lasers with the Rabi frequencies $\Omega_{\alpha}$ for the time $0\leq t_{1}\leq T$,
and the Laguerre-Gaussian dressing lasers with the Rabi frequencies
$\Omega_{\alpha}^{\mathrm{LG}}$ for the time $t_{2}=T-t_{1}$. Then,
for the stroboscopic frequency satisfying the condition $h,J\ll\omega_{\text{s}}\ll\Delta_{\pm}$,
the effective Rabi frequencies at the atomic positions are 
\begin{equation}
\Omega_{\alpha,k}^{\mathrm{eff}}=\frac{1}{T}(t_{1}\Omega_{\alpha}+t_{2}\Omega_{\alpha}^{\mathrm{LG}}e^{i\phi_{k,\alpha}})=\left|\Omega_{\alpha}^{\mathrm{eff}}\right|e^{i\phi_{k,\alpha}^{\mathrm{eff}}},\label{eq:Rabi effective}
\end{equation}
where 
\[
\left|\Omega_{\alpha}^{\mathrm{eff}}\right|=\frac{1}{T^{2}}[(t_{1}\Omega_{\alpha}+t_{2}\Omega_{\alpha}^{\mathrm{LG}}\cos\phi_{k,\alpha})^{2}+(t_{2}\Omega_{\alpha}^{\mathrm{LG}}\sin\phi_{k,\alpha})^{2}]^{1/2}
\]
and
\[
\tan\phi_{k,\alpha}^{\mathrm{eff}}=\frac{\sin\phi_{k,\alpha}}{\cos\phi_{k,\alpha}+t_{1}/t_{2}}.
\]
We see that, by varying the ratio $t_{1}/t_{2}$, one can generate
values $\phi_{k,\alpha}^{\mathrm{eff}}\in[0,\,\phi_{k,\alpha}]$ for
the effective phases. With the effective Rabi frequencies~\eqref{eq:Rabi effective},
the resulting phases of the flip-flop terms are not multiples of $2\pi$
anymore, and the corresponding Hamiltonian~\eqref{eq:Heff0-1} belongs
now to the unitary ensemble.

\section{Experimental considerations for SFF\label{sec:Appendix.Experimental-considerations}}

In this Appendix we discuss in more details the experimental challenges
of the SFF protocol, which limit the achievable system sizes. We also
present estimations of the coherence times and the available system
sizes for our Rydberg tweezer implementation.

\emph{Time scales:} – Propagation up to the Heisenberg time $\tau_{{\rm H}}\sim2^{L}/JL$
is limited by the finite coherence time of the quantum simulator.
First, the preparation time (with $M$ measurements) is $t_{\mathrm{prep}}\approx t_{0}2^{M}\leq\pi(JL)^{-1}2^{M-1}$
(see Sec.~\ref{subsec:Preparation MCE}). Therefore, the lower bound
on the coherence time is $t_{\mathrm{coh}}>t_{\mathrm{prep}}\approx(JL)^{-1}2^{M-1}$.
Observation of the behavior of $K(\tau)$ at times $\tau\sim\tau_{\mathrm{H}}$,
requires coherence times $t_{\mathrm{coh}}>\tau_{\mathrm{H}}\sim(JL)^{-1}2^{L}$.
In the case of a dominant individual single-spin decoherence with
the dephasing rate $\gamma_{{\rm d}}$, the corresponding condition
reads $J/\gamma_{{\rm d}}>2^{L}$. This limits observation of $K(\tau)$
at times $\tau\sim\tau_{{\rm H}}$ to moderate system sizes (see estimates
for the Rydberg tweezer array below). We note, however, that, even
when the Heisenberg time $\tau_{{\rm H}}$ is not accessible, the
transition to chaotic dynamics at the Thouless time $\tau_{{\rm Th}}$
and the distinct behaviors of the SFF for quantum chaotic {[}$K(\tau)\sim\tau${]}
and integrable systems {[}$K(\tau)\sim{\rm const}${]} takes place
at much shorter times to be compared with $t_{{\rm coh}}$. In general,
systems with smaller $L$ are characterized by shorter characteristic
times, and require fewer experimental runs to resolve the key features
(see discussion below). On the other hand, finite size effects tend
to wash out the characteristic features of $K(\tau)$ signaling the
chaotic behavior. 

\emph{Signal magnitude:} – The scaling of typical SFF values can be
estimated as $K_{\infty}\sim N_{\Delta E}^{-1}$ for the late time
plateau value and as $K(\tau\approx\tau_{{\rm Th}})\sim N_{\Delta E}^{-3/2}$
\citep{Cotler2017} for the SFF minimum preceding the Thouless time
$\tau_{{\rm Th}}$, here $N_{\Delta E}\gg1$ is the number of eigenstates
in the initial MC ensemble $\rho_{{\rm mc}}$. Preparation of MC ensemble
(see Sec.~\ref{subsec:Preparation MCE}) using $M$ filtering steps
(c-qubits) produces $\rho_{{\rm mc}}$ with $N_{\Delta E}\sim2^{L-M}$
eigenstates. Since the protocol has a success probability $p_{{\rm mc}}\sim2^{-M}$
and the QND measurement scheme can exploit the prepared state several
times ($N_{{\rm reuse}}$) one can perform $N$ measurements per data
point in $N_{{\rm run}}^{(1)}\simeq N/(p_{{\rm mc}}N_{{\rm reuse}})$
experimental runs per one disorder realization. On the other hand,
the threshold signal level which can be distinguished from the shot
noise after averaging over $N_{{\rm d}}$ realizations of disorder
in the spin Hamiltonian and $N$ measurements per one disorder is
given by $K_{*}\sim1/(N\sqrt{N_{{\rm d}}})$ (Appendix~\ref{sec:Appendix.Measurement noise}).
We find the necessary number of measurements using the condition $K_{*}\sim K_{\infty}$.
Thus, the number of experimental runs $N_{{\rm run}}=N_{{\rm d}}N_{{\rm run}}^{(1)}$
per data point necessary to resolve the features of interest in $K(\tau)$,
is given by $N_{{\rm run}}>2^{L}\sqrt{N_{{\rm d}}}/N_{{\rm reuse}}$.

The number of filtering steps $M$ does not affect $N_{{\rm run}}$
for the probabilistic preparation scheme, therefore, it is enough
to use $M\sim3$ to eliminate the contribution of spectral edges in
the SFF. However, it might be possible to use a semi-deterministic
preparation scheme, e.g., by populating excited energy eigenstates
in some energy interval by driving or quenching the system \citep{Senko2014}
followed by verification via $M$ filtering steps with high probability
($\sim1$) of success. In this case, the necessary number of experimental
runs per data point can be improved $N_{{\rm run}}>2^{L-M}$.

\emph{Rydberg tweezer implementation: –} We conclude with a discussion
of imperfections for the Rydberg tweezer implementation, in particular
decoherence rates and the effectiveness of the Rydberg blockade. We
also elaborate on geometrical limitations on interatomic distances
which ensure the validity of our Rydberg dressing and put a constraint
on the maximal size of the system.

We start with the discussion of the decoherence effects which limit
the duration of the experiment. They originate from the finite lifetime
of the Rydberg atomic states and from the error rate in the gate operation
caused by non-perfect Rydberg blockade mechanism (the term with $H_{\text{spin}}^{\prime}$
in \eqref{eq:H_system-qubit-1}). The former is characterized by two
dimensionless parameters $\kappa_{1}\equiv\gamma_{\text{d}}^{\prime}/\left|H_{\text{spin}}\right|\sim\gamma_{\text{d}}^{\prime}/(JL)$
and $\kappa_{2}\equiv\xi_{\pm}^{2}\gamma_{\text{d}}L/\left|H_{\text{spin}}\right|\sim\xi_{\pm}^{2}\gamma_{\text{d}}/J$,
for the control atom and the spin system, respectively. Here $\gamma_{\text{d}}^{\prime}$
and $\gamma_{\text{d}}$ are the spontaneous emission rates for the
corresponding Rydberg states, and for the system atoms we take into
account the collective enhancement ($\sim L$) of the spontaneous
emission rate due to highly entangled nature of the many-body excited
states. The factor $\xi_{\pm}^{2}$ in $\kappa_{2}$ represents the
admixture of the Rydberg state as a result of the dressing. In a similar
way, the error rate in the gate operation can be quantified by the
parameter $\kappa_{3}\equiv\left|H_{\text{spin}}^{\prime}\right|/\left|H_{\text{spin}}\right|$
which can be estimated {[}see Eqs.~\eqref{eq:Jxy} and \eqref{eq:Jz}{]}
as $\kappa_{3}\sim(R/R_{\text{b}})^{24}$ for $R<R_{\mathrm{b}},$
where $R$ is the distance between the control atom and the system
atoms and $R_{\text{b}}\equiv\sqrt[6]{\left|C_{6}^{\prime}/\Delta_{\pm}\right|}$
is the Rydberg blockade radius with $C_{6}^{\prime}$ being the interaction
constant {[}see Eq.~\eqref{eq:vdw_control-1}{]}. We note here the
very high power in the above estimate such that $\kappa_{3}$ decreases
very rapidly for $R<R_{\text{b}}$, giving, for example, $\kappa_{3}\sim10^{-3}$
for $R/R_{\text{b}}=0.8.$ The largest of these parameters $\kappa=\max\{\kappa_{i}\}$
sets the upper bound for the time during which the evolution is coherent
and follows the ideal QND Hamiltonian \eqref{eq:QNDHamiltonian},
$t_{\mathrm{coh}}\sim(JL\kappa)^{-1}$.

As an example let us consider the Rydberg states with $n^{\prime}=71$
(control atom) $n=60$ (system atom) and the following parameters
of the dressing scheme: $\Delta_{\pm}=-9\,\text{MHz}$ and $\xi_{\pm}=0.2$.
The decoherence rates for the atoms are $\gamma_{d}\approx2\pi\times318\text{Hz}$,
$\gamma_{d}^{\prime}\approx2\pi\times406\text{Hz}$. The parameters
characterizing the spontaneous emission are then $\kappa_{1}\approx4.4\cdot10^{-4},\kappa_{2}=1.6\cdot10^{-4}$
for $L\sim10$ atoms and the Rydberg blockade radius is $R_{\text{b}}\approx6.5\mu\text{m}$
(see Fig.~\ref{Fig4}). For $\kappa_{3}$ to be of the same order
or smaller, one should have $R\sim R_{\text{max}}=0.75R_{\text{b}}\approx5\mu\text{m}$
as an upper bound on the distance between the system and control atoms.
In our system $\kappa_{3}$ is the largest decoherence parameter and
it limits the coherence times to $t_{\mathrm{coh}}\sim10^{2}J^{-1}$.

The constraint on the distance between the system atoms is related
to the validity of our dressing scheme which is to say the validity
of the perturbative approach: It has to be larger than some minimal
value, $R_{ij}>r_{c}=2.4\mu\mathrm{m}$, see Appendix~\ref{p-s_interaction}.
Within our ring setup {[}see Fig.~\ref{fig:1}(b){]} in which the
radius is given by $R_{\text{max}}$ and the separation between system
atoms is limited by $r_{c}$, simple geometrical considerations give
the maximal number of system atoms which is with both constraints,
$L_{\text{max}}\simeq\pi/\text{arcsin\ensuremath{\left[r_{\text{c}}/\left(2R_{\text{max}}\right)\right]}}=12$,
the number used in our numerical simulations. 

\section{Preparation of a microcanonical ensemble\label{sec:Appendix.Preparation-of-MCE}}

In this Appendix we present detailed description of the preparation
of microcanonical ensembles via low resolution PEA as discussed in
Sec.~\ref{subsec:Preparation MCE}. As explained in the main text
the preparation procedure involves $M$ rounds of the QND interaction
entangling the spin system and the c-qubit according to $\mathcal{U}(t_{m},\delta)=\exp\left\{ -i[(H_{{\rm spin}}-\delta)\otimes\ket0\bra0]\,t_{m}\right\} $
followed by qubit measurements.

After the entanglement, the measurement of the c-qubit with a measurement
outcome $v_{m}=\{\pm\}$ collapses a state $\rho$ of the spin system
into a conditional (unnormalized) state
\[
\rho\to\mathcal{M}(v_{m},t_{m},\delta)\rho\mathcal{M}^{\dagger}(v_{m},t_{m},\delta),
\]
where the measurement operators $\mathcal{M}(\pm,t,\delta)=\bra{\pm}\mathcal{U}(t,\delta)\ket{+}$
act in the system Hilbert space and have a diagonal representation
in the energy eigenbasis 

\begin{align}
\mathcal{M}(+,t,\delta) & =\sum_{\ell}e^{-i(E_{\ell}-\delta)t/2}\cos[(E_{\ell}-\delta)t/2]\ket{\ell}\bra{\ell},\label{eq:App.M_op+}\\
\mathcal{M}(-,t,\delta) & =i\sum_{\ell}e^{-i(E_{\ell}-\delta)t/2}\sin[(E_{\ell}-\delta)t/2]\ket{\ell}\bra{\ell}.\label{eq:App.M_op-}
\end{align}
The full sequence of c-qubit measurements with outcomes $\vec{v}\equiv\{v_{m}\}$
defines the resulting unnormalized state of the spin system
\[
\tilde{\rho}_{{\rm out}}=\mathcal{\mathbb{M}}(\vec{v},\delta)\rho_{{\rm in}}\mathcal{\mathcal{\mathbb{M}}^{\dagger}}(\vec{v},\delta),
\]
where $\mathcal{\mathcal{\mathbb{M}}}(\vec{v},\delta)=\prod_{m=0}^{M-1}\mathcal{M}(v_{m},t_{m},\delta)$.
As a result, the unnormalized probability $p_{\ell}$ for the eigenstates
$\ket{\ell}$ with the eigenenergy $E_{\ell}$ to appear in $\rho_{{\rm out}}$
is 

\begin{align*}
p_{\ell}=\bra\ell\tilde{\rho}_{{\rm out}}\ket\ell & =P(\vec{v},E_{\ell}-\delta)\,\bra\ell\rho_{{\rm in}}\ket\ell,\\
P(\vec{v},x) & =\prod_{m=0}^{M-1}P_{m}(v_{m},x),
\end{align*}
where $P_{m}(\pm,x)=\left\{ 1\pm\cos(xt_{m})\right\} /2$.

If we now select a run with all readouts ``$+$'', $v_{m}=\{+\},\,m=0,\ldots,M-1$,
we obtain the output state with the narrow energy distribution

\begin{align}
 & p_{\ell}=P(+_{0},\ldots+_{M-1},E_{\ell}-\delta)\,\bra\ell\rho_{{\rm in}}\ket\ell,\label{eq:App.mc_filter}\\
 & P(+_{0},\ldots+_{M-1},x)=\left\{ \frac{\sin(2^{M}t_{0}x)}{2^{M}\sin(t_{0}x)}\right\} ^{2}.\label{eq:App.mc_filter2}
\end{align}
The success probability of the protocol is $p_{{\rm mc}}\equiv\sum_{\ell}p_{\ell}$.
The expressions~\eqref{eq:App.mc_filter} and~\eqref{eq:App.mc_filter2}
are used in Sec.~\ref{subsec:Preparation MCE}.

\section{Shot noise in SFF Measurements \label{sec:Appendix.Measurement noise}}

Here we determine the shot noise in the SFF measurement. As described
in Secs.~\ref{sec:Measurement-of-SFF-Hamiltonian} and~\ref{sec:Measurement-of-SFF-Floquet},
the measurement of the SFF involves the estimation of the expectation
values $\braket{\sigma^{x,y}}$ for the c-qubit by averaging results
of $N$ measurements. The statistical properties of such averaging
can be described by introducing $N$ copies of the c-qubit and the
spin system and considering the fluctuations of the collective spin
$S_{x,y}=N^{-1}\sum_{k=1}^{N}\sigma_{k}^{x,y}$ of the $N$ c-qubits.
In particular, we consider each copy of the c-qubit and spin system
to be initialized in the states $\ket+=\frac{1}{\sqrt{2}}(\ket{0}+\ket{1})$
and $\rho_{{\rm in}}$, respectively, and entangled via a controlled
unitary $\mathcal{U}(\tau)=U(\tau)\otimes\ket0\bra0+\mathbb{I}\otimes\ket1\bra1$.
In the case Hamiltonian dynamics we have $\mathcal{U}(\tau)=e^{-i\mathcal{H}_{{\rm QND}}\tau}$
and, consequently, $U(\tau)=e^{-iH_{{\rm spin}}\tau}$. Since we are
interested in the noise due to a finite number of measurements, in
what follows we consider a specific and fixed realization of disorder
in the Hamiltonian $H_{{\rm spin}}$. 

After the QND interaction, the state of the $N$ copies of the systems
reads

\[
\rho(\tau)=\left[\mathcal{U}(\tau)\rho_{{\rm in}}\otimes\ket+\bra+\mathcal{U}(\tau)^{\dagger}\right]^{\otimes N}.
\]
The consecutive measurements of the c-qubits yield an averaged outcome
which is given by an eigenvalue $m_{x,y}$ of the observable $S_{x,y}$.
The statistical distribution of the outcomes $m_{x,y}$ is characterized
by the corresponding moments of the collective spin observables with
respect to the state $\rho(\tau)$. 

The mean values read
\begin{align*}
\braket{S_{x}(\tau)} & =\frac{1}{N}\sum_{k=1}^{N}{\rm Tr}\left\{ \rho(\tau)\left(\mathbb{I}\otimes\ket{0}_{k}\!\bra{1}+\mathbb{I}\otimes\ket{1}_{k}\!\bra{0}\right)\right\} \\
 & =\frac{1}{2}{\rm Tr}\left\{ \rho_{{\rm in}}\left[U(\tau)+U(\tau)^{\dagger}\right]\right\} ={\rm Re}\braket{U(\tau)},\\
\braket{S_{y}(\tau)} & ={\rm Im}\braket{U(\tau)},
\end{align*}
where we expressed the operators $\sigma_{k}^{x,y}$ explicitly in
the basis $\ket{0(1)}_{k}$ of $k$th c-qubit. Thus, one can use the
first moments to evaluate the average of the unitary operator $\braket{U(\tau)}=\braket{S_{x}(\tau)}+i\braket{S_{y}(\tau)}\approx m_{x}+im_{y}$. 

The second and fourth moments of $S_{x}$ read 
\begin{align*}
\braket{S_{x}^{2}(\tau)} & =\frac{1}{N^{2}}\left\{ N(N-1)\left[{\rm Re}\braket{U(\tau)}\right]^{2}+N\right\} ,\\
\braket{S_{x}^{4}(\tau)} & =\frac{1}{N^{4}}\big\{ N(N-1)(N-2)(N-3)\left[{\rm Re}\braket{U(\tau)}\right]^{4}\\
+ & 2N(N-1)(3N-4)\left[{\rm Re}\braket{U(\tau)}\right]^{2}+N(3N-2)\big\}.
\end{align*}
This result and similar expressions for $S_{y}$ allows us to express
the SFF and its fluctuations (for a single disorder realization) as
\begin{align*}
K(\tau) & \equiv\left[{\rm Re}\braket{U(\tau)}\right]^{2}+\left[{\rm Im}\braket{U(\tau)}\right]^{2}\\
 & =\frac{N}{N-1}\left(\braket{S_{x}^{2}(\tau)}+\braket{S_{y}^{2}(\tau)}-\frac{2}{N}\right)\\
 & \approx m_{x}^{2}+m_{y}^{2}-\frac{2}{N}\pm\mathcal{O}\left(\frac{1}{N}\right).
\end{align*}
More precisely, the variance of the SFF estimation for $N\gg1$ reads
\begin{align*}
{\rm var}\left[K(\tau)\right] & =\Big\langle\left[S_{x}(\tau)^{2}+S_{y}(\tau)^{2}\right]^{2}\Big\rangle-\left[\braket{S_{x}^{2}(\tau)}+\braket{S_{y}^{2}(\tau)}\right]^{2}\\
 & \approx\frac{4}{N}K(\tau)+\frac{4}{N^{2}}.
\end{align*}
The signal-to-noise-ratio is thus given by $\smash{{\rm SNR}\equiv K(\tau)/\sqrt{{\rm var}\left[K(\tau)\right]}}$.
The SNR grows linearly $\mathcal{{\rm SNR}}\sim K(\tau)N/2$ with
$N$ up to $N\sim2(1+\sqrt{2})/K(\tau)$ where the SNR becomes 1.
For a given number $N$ of measurements and a fixed disorder, values
of the SFF above the threshold of $K_{*}^{(1)}\equiv2(1+\sqrt{2})/N$
can thus be determined with an SNR that is larger than 1. The threshold
value averaged over $N_{{\rm d}}$ realization of disorder $K_{*}=K_{*}^{(1)}/\sqrt{N_{{\rm d}}}$
is presented in Sec.~\ref{subsec:Experimental-challenges} and shown
as horizontal lines in Figs.~\ref{fig:SSF_L=00003D12} and \ref{fig:SFF_experiment_simulation}.
A further increase of the number of measurements results in a slower
growth with $\mathcal{{\rm SNR}}\sim\sqrt{K(\tau)N}/2$.

\section{Recycling of the microcanonical state in the SFF measurement\label{sec:Appendix.Recycling-of-the}}

In the Appendix we show that the SFF can be measured sequentially
at different times $\tau_{i}$ using a single initial state $\rho_{{\rm mc}}$.
Since the state $\rho_{{\rm mc}}$ is diagonal in the energy basis
it commutes with the QND Hamiltonian~\eqref{eq:QNDHamiltonian}.
Therefore, the state is not perturbed after averaging over measurement
results for a certain time $\tau$:
\begin{align*}
\rho_{{\rm out}} & =\mathcal{M}(+,\tau,\delta)\rho_{{\rm mc}}\mathcal{M}^{\dagger}(+,\tau,\delta)\\
 & +\mathcal{M}(-,\tau,\delta)\rho_{{\rm mc}}\mathcal{M}^{\dagger}(-,\tau,\delta)\\
 & =\rho_{{\rm mc}}.
\end{align*}
Here $\mathcal{M}(\pm,\tau,\delta)$ are the measurement operators
defined in Eqs.~\eqref{eq:App.M_op+} and \eqref{eq:App.M_op-}.

Consequently, one can recycle the prepared microcanonical sate as
long as the decoherence in the spin system is negligible $\sum_{i}\tau_{i}\ll t_{{\rm coh}}$.
Furthermore, the above applies to any initial state as only the diagonal
part of its density matrix contribute to the SFF. This result is used
in the Sec.~\ref{subsec:SFF_protocol_in Hamiltonian systems} and~\ref{subsec:Numerical-simulation-of-SFF}.

\section{Measurement of the Heisenberg time\label{sec:Appendix.Measurement-of-Heisenberg}}

In this Appendix, we show how the Heisenberg time $\tau_{{\rm H}}$
and the late-time plateau of the SFF $K_{\infty}$ can be obtained
from the probability $p_{{\rm mc}}$ with which the preparation of
a microcanonical ensemble as described in Sec.~\ref{subsec:Preparation MCE}
succeeds. Since the RMT form of the SFF is fully determined by these
two parameters, the method we describe in the following can be used
to validate the unbiased measurement of the SFF according to Sec.~\ref{sec:Measurement-of-SFF-Hamiltonian}
in the regime in which RMT is applicable.

The preparation scheme assumes that the spin system is initialized
in the infinite temperature state $\rho_{\infty}=\mathcal{D}^{-1}\sum_{\ell=1}^{\mathcal{D}}\ket\ell\bra\ell$,
where $\mathcal{D}$ is the Hilbert space dimension and $\ket\ell$
are the eigenstates of the spin Hamiltonian. A microcanonical ensemble
can then be prepared through $M$ successive projections of the c-qubit
to the state $\ket+$, which occur each time after it the c-qubit
has been entangled with the spin system. The success probability for
this procedure reads

\begin{align}
p_{{\rm mc}} & =\frac{1}{\mathcal{D}}\sum_{\ell}P_{+M}(E_{\ell}-\delta)\nonumber \\
 & \approx\frac{1}{\mathcal{D}}\frac{1}{t_{0}\delta_{E}}\int_{-1}^{1}P_{+M}(x)dx\equiv\frac{1}{\mathcal{D}}\frac{1}{t_{0}\delta_{E}}\mathcal{I}_{M},\label{eq:app_p_mc}
\end{align}
where the filter function $P_{+M}(x)$ is given by the Eq.~\eqref{eq:mc_filter}
and $\delta_{E}$ is the mean level spacing. Here we assume that the
filter function is narrow enough such that smooth changes in the density
of states can be neglected. For $M\gg1$, the integral $\mathcal{I}_{M}$
converges to $\pi2^{-M}$. Therefore, the mean level spacing can be
found as
\[
\delta_{E}\approx\frac{\mathcal{I}_{M}}{\mathcal{D}p_{{\rm mc}}}\frac{1}{t_{0}}\approx\frac{\pi2^{-M}}{\mathcal{D}p_{{\rm mc}}}\frac{1}{t_{0}},
\]
and the corresponding Heisenberg time reads
\begin{equation}
\tau_{{\rm H}}\equiv\frac{2\pi}{\delta_{E}}\approx2^{M+1}t_{0}\mathcal{D}p_{{\rm mc}}.\label{eq:app_t_H}
\end{equation}
The late-time plateau of the SFF can be estimated in a similar way.
It is given by 
\begin{align*}
K_{\infty} & =\sum_{\ell}f(E_{\ell})^{2}=\frac{1}{Z^{2}}\sum_{\ell}P_{+M}(E_{\ell}-\delta)^{2}\\
 & \approx\frac{1}{Z^{2}}\frac{1}{t_{0}\delta_{E}}\int_{-1}^{1}P_{+M}(x)^{2}dx\equiv\frac{\mathcal{S}_{M}}{Z^{2}}\frac{1}{t_{0}\delta_{E}}
\end{align*}
where $Z\equiv\sum_{\ell}P_{+M}(E_{\ell}-\delta)\approx\mathcal{I}_{M}/(t_{0}\delta_{E})$.
Taking into account that the integral $\mathcal{S}_{M}$ converges
to $\pi2^{1-M}/3$ for $M\gg1$ and using the expression for $p_{{\rm mc}}$
{[}Eq.~\eqref{eq:app_p_mc}{]} we obtain
\begin{equation}
K_{\infty}\approx\frac{\mathcal{S}_{M}}{\mathcal{I}_{M}^{2}}t_{0}\delta_{E}\approx\frac{\mathcal{S}_{M}}{\mathcal{I}_{M}}\frac{1}{\mathcal{D}p_{{\rm mc}}}\approx\frac{2}{3}\frac{1}{\mathcal{D}p_{{\rm mc}}}.\label{eq:app_K_inf}
\end{equation}
Equations \eqref{eq:app_t_H} and \eqref{eq:app_K_inf} are used in
the Sec.~\ref{subsec:Preparation MCE} to uniquely fix the RMT prediction
using experimental data.

\bibliography{Energy_measurement}

\end{document}